\documentclass[a4paper,12pt,twoside]{article}
\usepackage[utf8]{inputenc}
\usepackage[T1]{fontenc}
\usepackage{graphicx}
\usepackage{longtable}
\usepackage{wrapfig}
\usepackage{rotating}
\usepackage[normalem]{ulem}
\usepackage{amsmath}
\usepackage{amssymb}
\usepackage{capt-of}
\usepackage{hyperref}
\author{Shujiang Cao \and Shutao Cao\thanks{This version: August 27, 2025. Shujiang Cao: Renmin University of China, Beijing, China, caosj@ruc.edu.cn. Shutao Cao (corresponding author): Trent University, Peterborough, Canada, shutaocao@trentu.ca. The authors extend special thanks to Michael Veall for his support and encouragement throughout this work. The authors are grateful to Loretta Fung, Zhentong Lu, Pau Pujolas, Michael Veall, and participants at the 2024 Mini-Conference on Productivity at McMaster University and the Canadian Economic Association 2024 Annual Meeting, for thoughtful suggestions. The empirical part of this research was conducted at the University of Ottawa, a part of the Canadian Research Data Centre Network (CRDCN). This service is provided through the support of the Canada Foundation for Innovation, the Canadian Institutes of Health Research, the Social Sciences and Humanities Research Council, and Statistics Canada, and through the support of the University of Ottawa. The Productivity Partnership provided funding. All views expressed in this work are our own, and any remaining errors are solely our responsibility.}}
\usepackage[innermargin=2cm,outermargin=2cm,vmargin=2cm]{geometry}
\usepackage{amsfonts,amsthm,amsmath,amssymb,array,graphicx,indentfirst, epsfig,lscape, setspace, rotating,paralist}
\usepackage{color, xcolor,threeparttable}
\usepackage{natbib,fancyhdr,titlesec,fourier}
\titlespacing{\section}{0pt}{0.1\parskip}{-\parskip}
\titlespacing{\subsection}{0pt}{\parskip}{-\parskip}
\titlespacing{\subsubsection}{0pt}{\parskip}{-\parskip}
\usepackage{hyperref}
\hypersetup{colorlinks=true, linkcolor=red,citecolor=blue,filecolor=cyan, urlcolor=magenta}

\date{}
\title{Worker Quality, Matching and Productivity Slowdown}
\hypersetup{
 pdfauthor={s.t.cao},
 pdftitle={Worker Quality, Matching and Productivity Slowdown},
 pdfkeywords={},
 pdfsubject={},
 pdfcreator={Emacs 30.2 (Org mode 9.7.11)},
 pdflang={English}}
\begin{document}

\maketitle
\begin{abstract}
Measured aggregate productivity and the income share of top earners are strongly and positively correlated in the Canadian data. Productivity slowdown since the early 2000s was accompanied with a flattening income share of top earners. Motivated by these facts, we study the role of firms' top-paid workers and worker matching in accounting for the slowdown of measured total factor productivity. We first estimate total factor productivity for Canadian firms in the period of 2003-2015, taking into account the assortative matching between top workers and non-top workers. Measured total factor productivity consists of the Hicks-neutral technology and the quality of top workers. Our estimation suggests that measured aggregate total factor productivity declined from 2003 to 2015, in line with that estimated by the statistical agency. The decline of measured productivity is entirely accounted for by the declining quality of top workers, while the Hicks-neutral technology improved. Both the within-firm changes and the cross-firm reallocation of top-worker quality are important in contributing to the decline of overall top-worker quality. We also discuss possible causes of declines in the quality of top workers, e.g., the emigration of top talents as studied in recent literature.
\end{abstract}
\section{Introduction}
\label{sec:orgb4fc481}
Productivity remains a key indicator in measuring the performance of economies and businesses. The growth of measured productivity in advanced economies has stagnated since the early 2000s, despite much technological progress. This slowdown in productivity growth has been attributed to multiple factors, including weakened investment in physical capital, a deceleration of innovation and research productivity, the rise of intangible capital, and population aging, among others.\footnote{\cite{goldin2024} provides a comprehensive review of recent studies on productivity slowdown.}

In this paper, we study another source of productivity slowdown, namely the influences of top workers and the matching of workers on measured total factor productivity. Focusing on the Canadian economy, we are motivated by the following facts regarding productivity and top income inequality:\footnote{Statistics are calculated from Statistics Canada Tables 36100208 (for productivity) and 11100055 (for top income).}

\begin{enumerate}
\item Between 2000 and 2019, labor productivity grew at an average annual rate of 0.89 percent, significantly lower than the average of 1.71 percent per year over 1982-2000. The labor productivity slowdown is primarily driven by declines in measured total factor productivity.

\item The income share of top earners and productivity have a strong positive correlation. Over the years of 1982-2019, the correlation coefficient was 0.84 for top 1 percent earners and labor productivity, and 0.66 for top 1 percent earners and total factor productivity. Conversely, the correlation between the income share of bottom 50 percent earners and productivity was negative. Notably, since the early 2000s, the income share of top earners has plateaued and subsequently declined, coinciding with the slowdown in productivity.
\end{enumerate}

These observations suggest that there may be a tight relationship between top income inequality and productivity growth. The causal relationship between the two variables has not been extensively studied, yet there are a few exceptions. \cite{aghion2019} show that increased innovation (measured by the number of patents per capita) increases the income share of top earners. If increased innovation improves productivity, their model would predict a positive relationship between productivity and the income share of top earners. \cite{jones2018} developed a growth theory to explain the differences in top income inequality between the U.S. and other countries. In their model, if the creative destruction rate increases (say, by entrants), it can reduce the top income inequality.

We use the firm-level data to estimate the extent to which firms' top-paid workers and their matching with other employees may have contributed to the decline in measured total factor productivity since the early 2000s. We first characterize the optimal matching between top workers and other employees in a firm's production as in \cite{eeckhout2018}. Taking into account the optimal matching, we estimate the firm-level production function and obtain firm-level total factor productivity (Hicks-neutral technology) for Canadian firms. Using productivity estimates and measured top-worker quality at the firm level, we construct the measured total factor productivity that comprises two components: the level of Hicks-neutral technology and the contribution of top-worker quality.\footnote{Measured total factor productivity refers to factors that shift the output holding the factor of production (capital and labor) constant.} In our model, the top worker within each firm refers to the employee with the highest matched earnings who can be the top manager or chief scientist. This definition of top earners at the firm level differs from that obtained on the income distribution over the economy's labor force, but the two definitions overlap.

If the "output" function of a team of matched top workers and others exhibits a constant elasticity of substitution, and if the distributions of top workers and non-top workers are Pareto, we establish that the optimal assortative matching between top and non-top workers is positive and linear in worker quality. This simple functional form allows us to apply the proxy-variable approach to estimating the production function as in \cite{olley1996} and \cite{ackerberg2015}, which uses production inputs as proxy for unobserved productivity.

We implement the estimation using the Canadian firm-level data over the period 2003-2015 and find that the measured aggregate total factor productivity declined, in line with the trend of productivity measured by the statistical agency. Reallocation---measured as the covariance between measured productivity and the output shares of firms---fully accounts for the productivity declines, while the unweighted average of firm-level measured productivity increased from 2003 to 2015. This suggests that, relative to less productive firms, those with higher levels of measured productivity produce less over time, indicating that productive firms may invest less and hire fewer workers over time, a reallocation of production inputs towards less productive firms.

Furthermore, we find that decreases in measured productivity are entirely due to declines in the quality of top workers, while the estimated total factor productivity (Hicks-neutral technology) increased throughout the sample period. The contribution of top-worker quality to productivity slowdown is more pronounced in the years following 2008, which is consistent with the fact that the income share of top earners started to fall in 2008. Both the unweighted average of top-worker quality and the covariance between top-worker quality and output shares of firms have declined in the sample period, with the latter falling at a more accelerated rate. This implies that firms with higher quality top workers over time have produced less, relative to those with lower quality top workers.

In our estimation, the worker quality is measured as the person effects in the decomposition of matched earnings, estimated following \cite{abowd1999}.\footnote{Using person effects from the estimated wage decomposition to approximate worker quality (or skill) is also found in previous studies, for example, see \cite{iranzo2008}, \cite{fox2011}, \cite{bender2018}, and \cite{lochner2024}, among others.} It has two components: the person fixed effect and the effect related to the worker's age and sex. The falling fixed effect of top workers is the principal driver of declining top-worker quality. The age-sex effect has been declining steadily but at a slower pace, reflecting the gradual aging in the Canadian workforce.

The quality of non-top workers also declined from 2003 to 2015, but to a lesser extent when contrasted with top workers. This resulted in a narrowing quality gap between top workers and the average non-top workers, we call the gap the match efficiency. Thus, the declined in match efficiency, considered exogenous in our model, is a contributor to the productivity slowdown. The role of top workers in influencing measured productivity is equivalent to the roles played by the match efficiency and the quality of non-top workers.

Our findings imply that to boost productivity growth, enhancing the quality of top workers is crucial. Although this paper does not delve into the reasons behind the decline in top-worker quality since the early 2000s, potential factors include the emigration of top talents and shifts in top income taxation. In particular, if innovation is more likely conducted by top workers, policy initiatives aimed at fostering productivity growth should prioritize to retain top workers and provide appropriate incentives to spur innovation.

\paragraph{Related Literature.} The analysis in this paper is built upon two strands of literature. First, we model assortative matching between two types of workers in large firms by extending \cite{eeckhout2018}. They built a theory of assortative matching with endogenous firm sizes, unifying theories of firm size determined by the span of control and the sorting patterns between managers and workers. Our model focuses on the sorting patterns between top workers (who can be managers) and the rest of work force in a firm, while allowing for Hicks-neutral technology. This enables us to examine the roles of both Hicks-neutral technology and sorting patterns in firms' productivity dynamics. Earlier literature (\cite{gabaix2008} and \cite{tervio2008}) studies matching between CEO and firms to explain the growth of CEO compensation, where the firm size is exogenous. Second, we estimate the production function with the proxy-variable approach following \cite{ackerberg2015} which extends \cite{olley1996}. In our model setting, the existence of analytical form of the optimal matching function allows us to apply their methods with minimal modification.

Our production function, taking into account the optimal matching, can be considered to endogenize the total factor productivity. Recent literature endogenizes the total factor productivity by making it a function of managerial ability (\cite{guner2018}) or the number of managers (\cite{chen2023}). Both papers use general equilibrium models to quantify the role of managers in productivity growth. The difference is that in our model, the total factor productivity has two components, Hicks-neutral technology and a function of matched worker quality.

This paper contributes to the understanding of the role of worker composition and worker quality in productivity dispersion and dynamics. We measure the worker quality with the person effects in wage decomposition estimated following \cite{abowd1999} (AKM). In studying the links between firm productivity and wages, \cite{lochner2024} used the person effects from the AKM estimation for worker quality in their production function, which is log-supermodular in Hicks-neutral technology and worker quality. Though the authors also estimate production function, their model is based on matching with search frictions ignoring the span of control, while ours is based on assortative matching with endogenous firm size so that we can examine the importance of top workers. Our paper is closely related to \cite{bender2018} who studied the relationship among productivity, management practices, and worker quality. In their production function, the worker quality in firm, measured as the geometric average of fixed effects of managers and non-managers, is a component of total factor productivity, but they do not endogenize the matching between managers and non-managers. In our model, both total factor productivity (Hicks-neutral technology) and worker quality affect output. In addition, we estimate the production function (and productivity) that overcomes the simultaneity bias of ordinary least squares estimation as in \cite{olley1996} and subsequent studies.

Previous studies have found that workforce composition can affect firm productivity. \cite{iranzo2008} also use the worker fixed effects to approximate the worker quality (skill) and also estimate production function incorporating worker quality, but they do not model the matching between workers. Instead they estimate the role of within-firm dispersion of worker quality in output dynamics. Other studies use educational attainment level and worker characteristics to approximate human capital, see for example \cite{liu2010} and \cite{fox2011}. These authors do not model the matching between top workers and non-top workers.

Finally, this paper contributes to understanding the sources of productivity slowdown in Canada, often studied using the United States as a reference point. \cite{cao2017} extends Hulten's theorem regarding aggregate total factor productivity to a small open economy setting and finds that two sectors (commodity and machinery and equipment) account for most of the slowdown of total factor productivity since the early 2000s, which is further studied and confirmed by \cite{conesa2019}. The share of research and development in aggregate output was declining since the early 2000s, coinciding with productivity slowdown. \cite{ranasinghe2017} develops a heterogeneous-firm model to rationalize the slowdown of innovation as a source of stagnant productivity. None of these papers examines the link between worker quality and total factor productivity.

Our paper complements to \cite{macgee2025}. These authors document that the top income inequality alone accounts for the Canada-U.S. gap in labor productivity. Our paper provides a micro-level evidence on the link between top workers and productivity growth, and our findings are consistent with MacGee and Rodrigue's that are based on a parsimonious neoclassical growth model.

In the rest of the paper, we characterize the optimal matching between top workers and non-top workers in Section 2. In Section 3, we measure the worker quality by estimating the two-way fixed effect equation of wage decomposition. We then estimate the production function and obtain measures of productivity in Section 4. With the estimated production function, we analyze the dynamics of measured aggregate productivity and cross-firm productivity dispersion in Section 5. In Section 6, we discuss the sources of declining top-worker quality. Finally, we conclude in Section 7 and provide thoughts for further research.
\section{Production with assortative matching}
\label{sec:org3dc45df}
\subsection{Optimal production}
\label{sec:orgae8aad2}
Our focus is on the optimal team formation and its implications for productivity. We extend \cite{eeckhout2018} by introducing Hicks-neutral technology (total factor productivity) and the match efficiency in production. A firm with a level \(\omega\) of Hicks-neutral technology hires a team with two types of members. The team consists of one worker with type or quality \(y\) and \(l(x)\) workers with type or quality \(x\).\footnote{In this paper, terms of worker type, worker quality and worker ability are interchangeable.} By assumption, \(y\ge x\). The type-\(y\) worker can be the chief manager, or the lead scientist or lead engineer who organizes and coordinates all tasks in the production. Type-\(x\) workers are employees holding supporting positions or working on various specific tasks. A firm's production function is given by
\[f(\omega,x,y,l) = e^{\omega}\left[\alpha_x (e^{\omega_x} x)^{1-\sigma} + \alpha_y y^{1-\sigma}\right]^{\frac{\theta}{1-\sigma}} \cdot [l(x)]^{\alpha_l}, \]
where \(\omega\) represents the Hicks-neutral productivity, and \(1/\sigma\) is the elasticity of substitution between worker type \(x\) and top worker type \(y\).

We assume that the match between worker types \(x\) and \(y\) displays heterogeneity across firms, captured by \(\omega_x\), an exogenous match efficiency. If the value of \(\omega_x\) differs between two firms, it means that a team of type \((x,y)\) may produce different amounts of output when working for two different firms, even if the Hicks-neutral technology is equal between the two firms. In this sense, \(\omega_x\) is also a measure of mismatch. Imposing \(\omega_x\) breaks the monotonicity of matches. A top worker \(y\) may match with different types of non-top workers, depending on draws of the match efficiency.

The type-\(\omega\) firm's problem is to find a team that maximizes profits
\[\max_{\{x,y,l(x)\}} f(\omega,x,y,l(x)) - w(x) l(x) - r(y).\]

The first-order necessary conditions with respect to \(y\), \(x\), and \(l\) are respectively given by
\[f_y - r'(y) =0, \ \  f_x - w'(x) l(x) = 0, \ \  f_l - w(x) =0.\]

The optimal matching between worker types should also satisfy the market clearing conditions. Let the optimal matching function be \(y = T(x)\), and let \(g(x)\) and \(h(y)\) respectively be the probability density functions of types \(x\) and \(y\). If the match is positively assortative, the market clearing condition is given by
\[\int_{T(x)}^{\overline{y}} l(s) h(s) d s = \int_x^{\overline{x}} g(s) ds,\]
where the left-hand side is the aggregate quantity demanded for labor of type \(x\) or higher, and the right hand side is the aggregate quantity supplied of labor of type \(x\) or higher.

If the match is negatively assortative, the market clearing condition is given by
\[\int_{T(x)}^{\overline{y}} l(s) h(s) d s = \int_{\underline{x}}^{x} g(s) ds.\]

The equilibrium match between \(x\) and \(y\) can be solved for from firms' optimal demand for labor and the market clearing condition. With positive assortative matching (PAM), these conditions are differential equations as follows
\[f_x - w'(x) l(x) = 0; \quad f_l - w(x) =0; \quad T'(x) = \frac{\mathcal{H}(x)}{l(x)}.\]
The optimal condition regarding \(T'(x)\) is obtained from the market clearing condition.\footnote{For a mathematical treatment, see for example \cite{Maggi_2023}.} Here, \(\mathcal{H}(x) = g(x) / h(T(x))\).

With negative assortative matching (NAM), optimal conditions are a system of three differential equations as follows
\[f_x - w'(x) l(x) = 0; \quad f_l - w(x) =0; \quad T'(x) = -\frac{\mathcal{H}(x)}{l(x)}.\]

We focus on the positive assortative matching, as it is consistent with the data we use in which measured worker types \(x\) and \(y\) are positively correlated. Positive assortative matching requires the firm's production function to satisfy \((\theta > 0, \sigma\ge 1)\) or \((\theta <0, \sigma \le 1)\), as derived in Appendix \ref{sec:org1484cd0}. In addition, our primary focus is on the optimal matching of worker types, and we only need to recognize the endogeneity of the choice on \(y\) for estimation. Solving for y is unnecessary.
\subsection{Optimal matching under Pareto distributions}
\label{sec:org86de851}
Our goal is to bring the model to data and quantify the importance of match efficiency in productivity dynamics. To facilitate that, we would want the analytical form of the matching function. We thus impose parametric functional forms on worker distributions by assuming that \(x\) and \(y\) follow Pareto distributions.

With the Pareto distributions of worker types, the optimal match has an explicit functional form. Two additional considerations also help justify the use of Pareto distributions. First, assuming that worker types have Pareto distributions leads to the Cobb-Douglas form of aggregate production function, as shown in \cite{houtakker1955}, \cite{jones2005}, and \cite{lagos2006}. This result allows the aggregated productivity to be comparable with the productivity measured using an aggregate production function that displays constant returns to scale. Second, the Pareto assumption on the distribution of top-worker types is strongly supported by the data, and top-worker types display a Pareto distribution shape beyond a small threshold value of the type. For the non-top workers, the Pareto assumption on the distribution is also supported by the data but to a lesser extent.

Let the probability distribution of \(x\) be \(G(x) = 1 - P(X > x) = 1- \left(\underline{x}/x\right)^{\lambda_x}\) for \(x\ge \underline{x}\) and \(\lambda_x>0\). The probability density function for \(x\) is then \(g(x) = \lambda_x \underline{x}^{\lambda_x} / x^{\lambda_x+1}\). Let the probability distribution of \(y\) be \(H(y) = 1 - P(Y > y) = 1- \left(\underline{y}/y\right)^{\lambda_y}\) for \(y\ge \underline{y}\) and \(\lambda_y>0\). The probability density function for \(y\) is then \(h(y) = \lambda_y \underline{y}^{\lambda_y} / y^{\lambda_y+1}\). Given these functions, \(\mathcal{H}(x) = g(x) / h(T(x)) = C \cdot [T(x)]^{\lambda_y+1}/x^{\lambda_x+1}\) with \(C = \lambda_x \underline{x}^{\lambda_x}/(\lambda_y \underline{y}^{\lambda_y})\).

In Appendix \ref{sec:org1484cd0}, we derive the closed-form solution of the optimal matching function and wage rate of non-top workers. We show that if the production function displays a constant elasticity of substitution between \(x\) and \(y\) and distributions of worker types are Pareto, the optimal matching function is given by
\[y = T(x) = \Psi^{\frac{1}{1-\sigma}} e^{\omega_x} x,\]
with \(\Psi = \frac{\alpha_x (1-\alpha_l) [\theta + \alpha_1 (\lambda_y - \lambda_x)]}{\alpha_y \alpha_l [\theta- (1-\alpha_l)(\lambda_y + \lambda_x)]}\).

The equilibrium matching is linear in worker types. The shape of the matching function does not depend on the elasticity of substitution between top worker and non-top workers. The following condition is sufficient for PAM. For any \(\sigma\neq 1\),

\(\theta > \max \left\{\alpha_l(\lambda_x-\lambda_y), -(1-\alpha_l)(\lambda_x-\lambda_y)\right\}\) or
\(\theta < \min \left\{\alpha_l(\lambda_x-\lambda_y), -(1-\alpha_l)(\lambda_x-\lambda_y)\right\}\).

The optimal demand for labor is given by
\[l(x) = \Psi^{\frac{\lambda_y}{1-\sigma}} C \cdot e^{\omega_x \lambda_y} x^{\lambda_y - \lambda_x}.\]

The number of non-top workers may increase or decrease with the worker type, depending on the relative size of Pareto exponents \(\lambda_x\) and \(\lambda_y\). Both parameters determine the probability of having more extreme types. The smaller the parameter values, the larger the probability of having extremely high worker types. If the dispersion of top-worker types is smaller than that of non-top workers, the number of workers will be smaller for larger values of types of non-top workers.

Using the optimal condition \(f_l - w(x) = 0\), we find that the optimal wage rate for type \(x\) is
\[w(x) = \Lambda \cdot e^{\omega} \cdot e^{\omega_x(\theta - \lambda_y(1-\alpha_1))} \cdot x^{\theta + (1-\alpha_l)(\lambda_x-\lambda_y)},\]
where \(\Lambda\) is positive and consists of parameters only. The wage rate of non-top workers increases with the type if the assortative matching is positive.

Taking into account the optimal match, the production function becomes
\[f(\omega, x,T(x),l) = (\alpha_x + \alpha_y \Psi)^{\frac{\theta}{1-\sigma}} e^{\omega} (e^{\omega_x}x)^{\theta} [l(x)]^{\alpha_l}.\]
This production function does not identify output elasticity because both Hicks-neutral productivity \(\omega\) and the match efficiency \(\omega_x\) affect the firm's optimal choice on non-top worker type \(x\) and the work force, and both efficiency indexes are unobserved. In the approach to estimation we take below, the presence of two unobserved shocks prevents us from identifying the output elasticity of inputs. Alternatively, the production function can be transformed to the following form
\[f\left(T^{-1}(y),y,l\right) =\left(\alpha_x \Psi^{-1}+ \alpha_y\right)^{\frac{\theta}{1-\sigma}} e^{\omega} y^{\theta} \left[l\left(T^{-1}(y)\right)\right]^{\alpha_l},\]
which will be used to estimate the effect of top workers on aggregate productivity growth.
\section{Measuring the worker quality}
\label{sec:orgbb5df9d}
Before estimating the production function, we first need to measure worker quality, for which we use the person effects estimated from the two-sided fixed-effect equation of matched earnings following \cite{abowd1999} (henceforth AKM). We then label the estimated person effects of top workers as their quality, and the person effects of non-top workers as the quality of non-top workers. The wage equation implied by the optimal matching as obtained above shows that the wage rate (in logarithm) is determined by the Hicks-neutral technology, match efficiency, and the worker quality. These factors are additive and separable, hence the fixed effects in the AKM estimation are appropriate for measuring the worker quality.

For the AKM estimation, we use the Canadian Employer Employee Dynamics database (CEEDD) 2003-2015, which is a data infrastructure of the population of persons and businesses in Canada. Businesses and employees are linked through the dataset named the Report of Employment (ROE). Details of the CEEDD are described in Appendix \ref{sec:orge955ae1}. Annual matched earnings for each match, but not the wage rate, is available in the data. We therefore use the matched earnings for estimation.

The logarithm of matched earnings of employee \(i\) at firm \(j\) in year \(t\) can be decomposed into the firm fixed effect, the person fixed effect, effects of worker characteristics, and the residual component, as follows
\[\ln w_{it} = \beta_0 + \alpha_i + X_{it} \pmb{\beta} + \psi_{j(i,t)} + \varepsilon_{ijt},\] where \(\alpha_i\) is the worker fixed effect and \(\psi_{j(i,t)}\) is the firm fixed effect. Vector \(X_{it}\) includes quadratic and cubic terms in worker age interacted with a dummy variable for male. We normalize age by subtracting 40 from all ages, restricting the age profile to be flat at age 40 as in \cite{card2013}. In addition, we include a two-year dummy variable in the regression. Details of the AKM estimation are reported in Appendix \ref{sec:org1795b7d}.

In estimating the earnings equation, we make sure that the top-paid worker in a firm is not an owner of the firm. In the connected data set, at least one employee of a firm is also an owner of the firm in about 42 percent of firm-year observations. Further, an owner of a firm is also the top-paid employee of the firm in about 34 percent of firm-year observations. To properly estimate the quality of top-paid workers, in the AKM estimation, we drop all employee-year observations in which an employee also owns at least 1 percent of the firm in that year. The presence of employee owners makes it harder to interpret the decomposition of matched labor earnings.

We label the term \(h_{it} = \widehat{\alpha}_i + X_{it} \pmb{\widehat{\beta}}\) as the type (quality) of worker \(i\), which is the estimate of the fixed effect and effects associated with age and sex. \(h_{it}\) is the quality of top worker \(i\) in a firm if worker \(i\) is the top paid employee in the firm. For the type of non-top workers in each firm, we use the average of the fixed effects and effects associated with age and sex over all non-top workers that the firm hires.

Important facts stand out regarding matched earnings and worker quality estimates. For all employees in each firm, we calculate the earnings ratio between the top worker and each non-top worker. First, the average of the matched earnings ratios between the top worker and non-top workers across all employees is 2.62, and the median ratio is 2.07. Thus, on average, the top worker earns more than twice non-top workers with the same employer earn. Second, the average of top to non-top earnings ratios increases with the firm size. This ratio is about 2.25 among firms with fewer than 10 employees, and is 15 among firms with 500 employees or more. Third, across firms, matched earnings of top workers increases with the firm size. The median earnings of top workers in firms with fewer than 10 employees is about 12 times that in firms with 500 employees or more.

\begin{figure}[htbp]
\centering
\includegraphics[angle=0,scale=0.30]{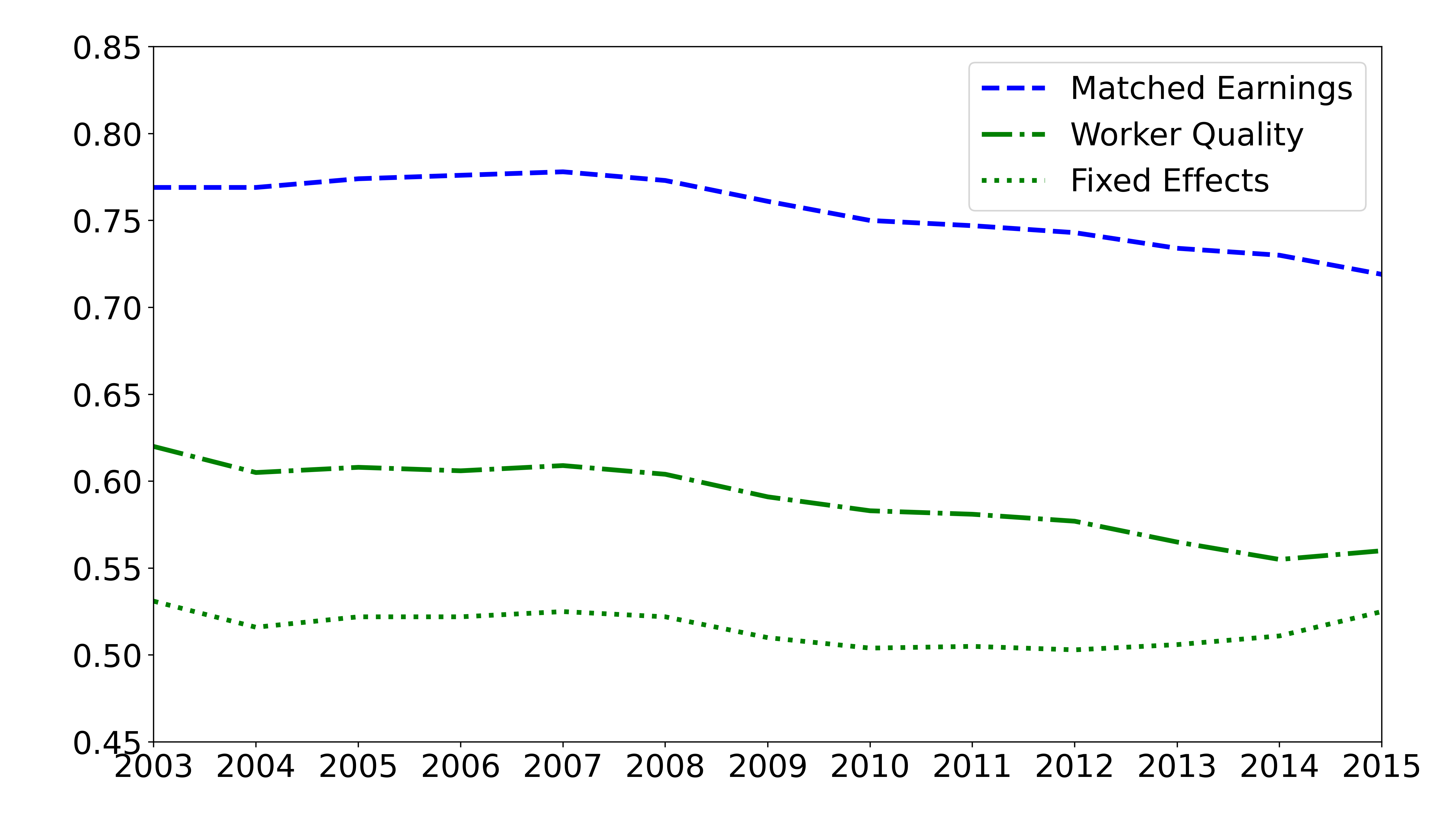}
\caption{\label{fig:org1b354f2}Median log-difference between top workers and non-top workers}
\end{figure}

For each firm, we also calculate the ratio of top worker quality relative to the \emph{average} of the quality of non-top workers in the same firm, which corresponds to \(y/x\) in the production function. We define similar ratios for matched earnings and the fixed effects as well. According to our estimation, the median of this ratio is 0.588 in logarithm. It means that, in the distribution of the quality of ratios over all firms, at the median the top worker's ability is 1.8 times the average ability of non-top workers. The median ratio exhibits a secular declining trend from 2003 to 2015, as shown by the dash-dotted line in Figure \ref{fig:org1b354f2} (values displayed are the logarithm of median ratios), which is mainly due to the declines of top-worker quality. We will examine these patterns in the next section.

\section{Measuring firm-level productivity}
\label{sec:orgeb66547}
\subsection{Estimation of production function}
\label{sec:org457d916}
The production function, taken into account optimal matches, can be transformed to the form given by
\[f\left(\omega, T^{-1}(y),y,l\right) =\left(\alpha_x \Psi^{-1}+ \alpha_y\right)^{\frac{\theta}{1-\sigma}} e^{\omega} y^{\theta} \left[l\left(T^{-1}(y)\right)\right]^{\alpha_l}.\]
We estimate output elasticity based on this production function and recover productivity \(\omega\) from the estimation. As noted in the above, the production function
\[f(\omega, x,T(x),l) = (\alpha_x + \alpha_y \Psi)^{\frac{\theta}{1-\sigma}} e^{\omega} (e^{\omega_x}x)^{\theta} [l(x)]^{\alpha_l},\]
does not identify output elasticity, because both Hicks-neutral productivity \(\omega\) and the match efficiency \(\omega_x\) affect the firm's optimal choice on non-top worker type \(x\) and the work force, violating the condition of single exogenous shock used in estimation. Both efficiency indexes are unobserved, so we cannot recover productivity from this production function.

We augment the production function to include physical capital as factor of production, and assume that the contribution of labor input and capital takes a Cobb-Douglas form. Capital is pre-determined in production in the current period. Allowing for capital in production does not alter the matching function in equilibrium.\footnote{Capital is not a choice in the current period. To see that adding capital does not alter the matching function, simply put capital (with power to \(\alpha_k\)) together with the productivity \(\omega\) and take them as one variable.}

The production function of firm \(j\) in time period \(t\) is
\[f\left(T^{-1}(y_{jt}),y_{jt},l_{jt}, k_{jt}\right) =\left(\alpha_x \Psi^{-1}+ \alpha_y\right)^{\frac{\theta}{1-\sigma}} e^{\omega_{jt}} y_{jt}^{\theta} \left[l\left(T^{-1}(y_{jt})\right)\right]^{\alpha_l} k_{jt}^{\alpha_k}.\]
In natural logarithm, the equation is written as
\begin{equation}
\label{eq:PF}
\ln f_{jt} = \beta_0 + \theta \ln y_{jt} + \alpha_l \ln l_{jt} + \alpha_k \ln k_{jt} + \omega_{jt} + \varepsilon_{jt},
\end{equation}
where we allow for measurement errors and unexpected shocks to production, \(\varepsilon_{jt}\), with \(\varepsilon_{jt} \sim \mathrm{N}(0,\sigma_u^2)\). The optimal demand for labor and the optimal match do not depend on the measurement error \(\varepsilon_{jt}\), and \(\omega_{jt}\) is uncorrelated with \(\varepsilon_{jt}\). We assume that the Hicks-neutral productivity evolves over time with an AR(1) process, \(\omega_{jt} = \rho\omega_{jt-1} + \xi_{jt}\). The innovation term \(\xi_{jt}\) is thus uncorrelated with the measurement error \(\varepsilon_{jt}\).

We estimate the production function (\ref{eq:PF}) using the proxy-variable approach by \cite{ackerberg2015}, which is built on \cite{olley1996}, among others. This approach overcomes the simultaneity bias of estimation by the ordinary least squares. We use the firm's choice of intermediate inputs as the proxy variable, \(m_{jt} = M(\omega_{jt}, y_{jt}, l_{jt}, k_{jt}, p_{gt},p_{mt})\). Here, \(p_{gt}\) and \(p_{mt}\) are respectively the price of gross output and the price of intermediate inputs, both at 3-digit NAICS level. The inverse of proxy variable is \(\omega_{jt} = M^{-1}(m_{jt}, y_{jt}, l_{jt}, k_{jt}, p_{gt},p_{mt})\). Substituting it for \(\omega_{jt}\), the production function becomes
\begin{equation}
\label{eq:OPstage1}
\ln f_{jt} = \phi (m_{jt}, y_{jt}, l_{jt}, k_{jt}, p_{gt},p_{mt}) + \varepsilon_{jt},
\end{equation}
where \(\phi (m_{jt}, y_{jt}, l_{jt}, k_{jt}, p_{gt},p_{mt}) = \beta_0 + \theta \ln y_{jt} + \alpha_l \ln l_{jt} + \alpha_k \ln k_{jt} + M^{-1}(m_{jt}, y_{jt}, l_{jt}, k_{jt}, p_{gt},p_{mt})\), it is nonlinear and may have no closed form. We label function \(\phi(\cdot)\) as \(\phi_{jt}\), as it captures the variation of (the logarithm of) output explained by inputs and prices. Taken into account \(\omega_{jt} = \rho\omega_{jt-1} + \xi_{jt}\), the production function can be written as
\begin{align}
& \ln f_{jt} = \beta_0 + \theta \ln y_{jt} + \alpha_l \ln l_{jt} + \alpha_k \ln k_{jt} \nonumber \\
& \quad \quad \quad + \rho \left[\phi_{jt-1} - (\beta_0 + \theta \ln y_{jt-1} + \alpha_l \ln l_{jt-1} + \alpha_k \ln k_{jt-1}) \right] + \xi_{jt} + \varepsilon_{jt}.
\label{eq:OPstage2}
\end{align}
Included in the square brackets is \(\omega_{jt-1}\).

The estimation takes two stages. In the first stage, we estimate Equation (\ref{eq:OPstage1}) by the ordinary least squares (OLS), where we approximate \(\phi_{jt}\) with a 3rd-order polynomial of \(\ln y_{jt}, \ln l_{jt}\), \(\ln k_{jt}\), \(\ln m_{jt}\) and the logarithm of prices. From the first stage, we obtain the estimate \(\widehat{\phi} (m_{jt}, y_{jt}, l_{jt}, k_{jt}, p_{gt},p_{mt})\), the variation of output explained by observed inputs and prices. In the second stage, we substitute \(\widehat{\phi}_{jt-1}\) for \(\phi_{jt-1}\) in Equation (\ref{eq:OPstage2}), and estimate the equation with the generalized method of moments (GMM). Because the labor input and the top-worker type are correlated with \(\xi_{jt}\), in estimating Equation (\(\ref{eq:OPstage2}\)) we use \(\mathbf{I}_{jt-1} = (1, \widehat{\phi}_{jt-1}, \ln m_{jt-1}, \ln y_{jt-1}, \ln l_{jt-1}, \ln k_{jt})\) as instrument variables. The moment conditions are given by
\[E\left[(\xi_{jt}+\varepsilon_{jt})\otimes \mathbf{I}_{jt-1}\right] = 0.\]

The second-stage estimation gives the estimates of output elasticity and the coefficient for serial correlation of productivity, as reported in Table \ref{tbl:PFest}. Estimates are statistically significant at 1 percent level for the majority of parameters and sectors. The estimated production function displays increasing returns to scale in all inputs and decreasing returns to scale in capital and labor. Output elasticity with respect to labor and capital are fairly stable across sectors. Overall, coefficient estimates appear reasonable.

We point out two issues in the estimation. First, we are unable to identify the coefficient of elasticity of substitution \(\sigma\). Under the CES form of "output" between worker types and the Pareto assumption of the distributions of worker types, the matching function is linear in worker types \(x\). The economically meaningful case is positive assortative matching, as is supported by the strong and positive correlation of types between the top workers and non-top workers found in data. Second, our estimation does not overcome the possible selection bias due to ignoring the endogenous exit decisions of firms, and this is mainly due to data quality. Many firms in the data set have about six years of data, but no information regarding firm exit is provided. The difficulty of taking into account firm exit in estimation lies in the lack of information regarding the exact time at which a firm exits. Most firms are very small, and measures of inputs or outputs may be significantly noisy or missing. A firm may still exist even though information on inputs and outputs is missing. This makes it challenging to estimate the exit probability since missing inputs and outputs may be unrelated to productivity.

\begin{landscape}
\begin{table}
\label{tbl:PFest}
\begin{threeparttable}[b]
\caption{Coefficients of production Function}
\centering
\small
\begin{tabular}{l|rr|rr|rr|rr|rr|r}
\hline
Sector (2-digit Naics) & \multicolumn{2}{c}{\(\beta_0\)} & \multicolumn{2}{c}{\(\theta\)} & \multicolumn{2}{c}{\(\alpha_l\)} & \multicolumn{2}{c}{\(\alpha_k\)} & \multicolumn{2}{c|}{\(\rho\)} & Observations \\
\hline
& Value & Stdder & Value & Stdder & Value & Stdder & Value & Stdder & Value & Stdder & \\
\hline
Mining (21) & 10.115 & 0.398 & 0.153 & 0.266 & 0.721 & 0.15 & 0.156 & 0.034 & 0.908 & 0.06 & 12980\\
Construction (23) & 10.987 & 0.113 & 0.417 & 0.051 & 0.777 & 0.019 & 0.079 & 0.009 & 0.702 & 0.007 & 381250\\
Manufacturing (31-33) & 11.654 & 0.329 & 0.935 & 0.175 & 0.578 & 0.051 & 0.043 & 0.026 & 0.808 & 0.012 & 287695\\
Wholesale trade (41) & 11.941 & 0.264 & 1.275 & 0.156 & 0.53 & 0.06 & 0.031 & 0.021 & 0.847 & 0.009 & 225985 \\
Retail trade (44-45) & 11.282 & 0.173 & 0.278 & 0.057 & 0.8 & 0.032 & 0.049 & 0.016 & 0.751 & 0.01 & 433260 \\
Transportation (48-49) & 10.467 & 0.093 & 0.448 & 0.123 & 0.633 & 0.067 & 0.145 & 0.01 & 0.785 & 0.037 & 61315 \\
Information (51) & 10.537 & 0.165 & 0.429 & 0.173 & 0.677 & 0.087 & 0.131 & 0.018 & 0.814 & 0.029 & 22955 \\
Finance, real estate (52-53) & 11.79 & 0.109 & 0.579 & 0.164 & 0.781 & 0.12 & 0.013 & 0.018 & 0.818 & 0.035 & 39000 \\
Professional (54) & 11.479 & 0.064 & 0.407 & 0.041 & 0.72 & 0.022 & 0.055 & 0.006 & 0.708 & 0.017 & 106325 \\
Management (55) & 10.552 & 0.253 & 0.384 & 0.2 & 0.705 & 0.193 & 0.118 & 0.051 & 0.794 & 0.087 & 8785 \\
Administration (56) & 10.714 & 0.047 & 0.521 & 0.074 & 0.601 & 0.064 & 0.113 & 0.006 & 0.701 & 0.034 & 94000 \\
Education (61) & 10.861 & 0.222 & 0.391 & 0.181 & 0.663 & 0.119 & 0.104 & 0.025 & 0.785 & 0.079 & 11420 \\
Health care (62) & 11.225 & 0.233 & 0.803 & 0.185 & 0.435 & 0.163 & 0.118 & 0.031 & 0.807 & 0.049 & 44885 \\
Arts, entertainment (71) & 11.167 & 0.129 & 0.188 & 0.132 & 0.768 & 0.106 & 0.04 & 0.038 & 0.71 & 0.076 & 31740 \\
Accommodation (72) & 10.509 & 0.114 & 0.183 & 0.039 & 0.761 & 0.021 & 0.089 & 0.011 & 0.633 & 0.009 & 298945 \\
Other services (81) & 10.768 & 0.057 & 0.501 & 0.04 & 0.659 & 0.023 & 0.103 & 0.005 & 0.703 & 0.011 & 171470 \\
\hline
\end{tabular}
\begin{tablenotes}
      \footnotesize
      \item Notes: Coefficients are estimated with the two-stage proxy variable approach. Standard errors are obtained via bootstrap to correct for the second stage using estimates from the first stage. Estimates are significant at 1 percent level for most parameters.
    \end{tablenotes}
\end{threeparttable}
\end{table}
\end{landscape}
\subsection{Estimating the match efficiency}
\label{sec:orgf17aede}
The measure of match efficiency between top workers and the rest cannot be recovered from the estimated production function. We thus bring the matching function to the data to obtain this measure. In natural logarithm, the matching function is
\[\ln y_{jt} = b_0 + \omega_{xjt} + \ln x_{jt},\]
where \(b_0 = \frac{1}{1-\sigma} \ln \Psi\). We assume that the match efficiency follows an AR(1) process, \(\omega_{xjt} = \rho_x \omega_{xjt-1} + u_{jt}\). We use the generalized method of moments to estimate the following equation
\[(\ln y_{jt} - \ln x_{jt}) = (1-\rho_x) b_0 + \rho_x(\ln y_{jt-1} - \ln x_{jt-1}) + u_{jt},\]
using \((\ln y_{jt-1} - \ln x_{jt-1})\) as the instrument variable. We implement the estimation by sector since parameter \(b_0\) involves sector-specific output elasticity.

\begin{figure}[htbp]
\centering
\includegraphics[angle=0,scale=0.30]{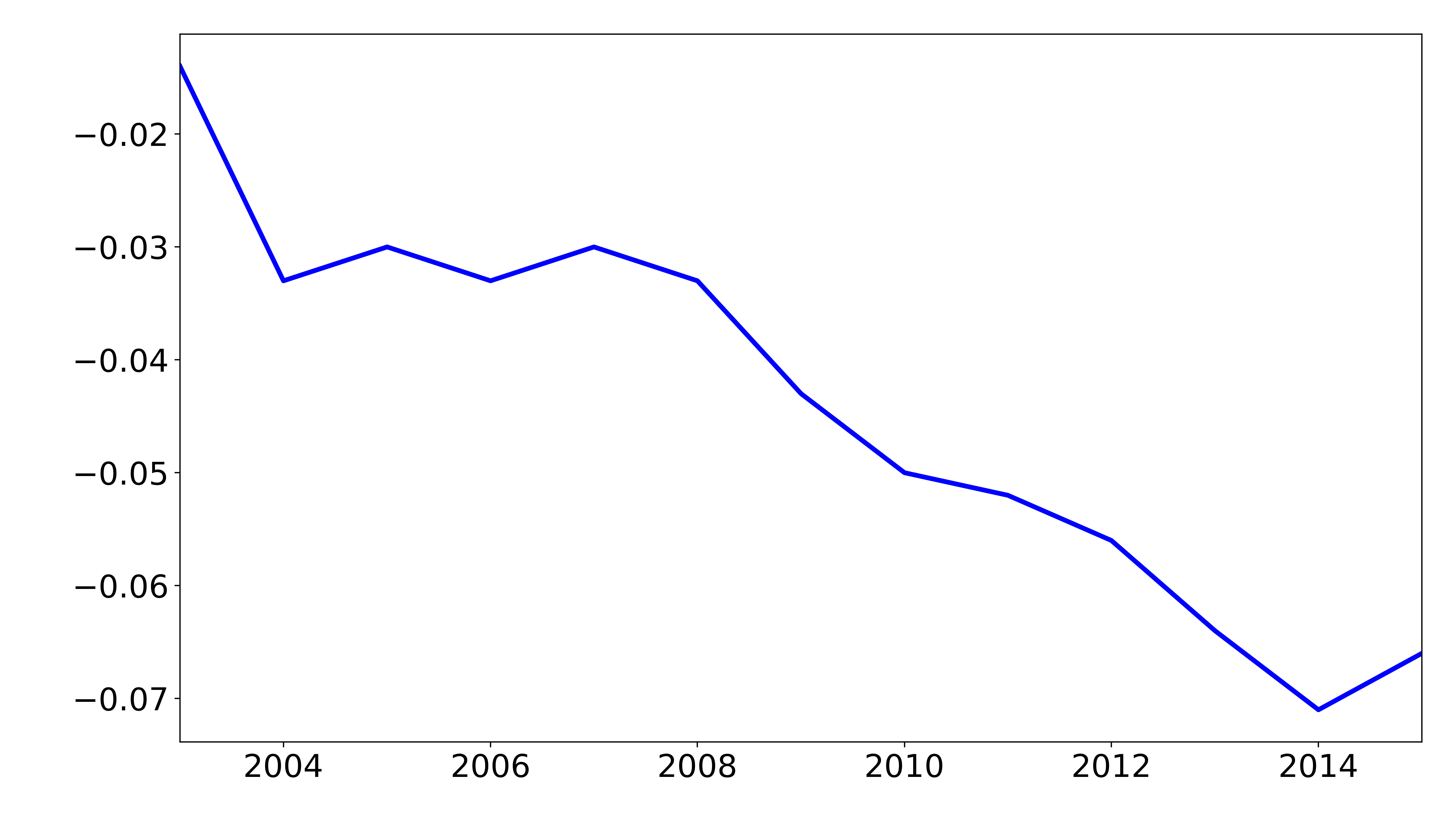}
\caption{\label{fig:org667477a}Median of match efficiency in logarithm}
\end{figure}

Estimates of the coefficient for serial correlation of match efficiency range from 0.6 to 0.8. The median of match efficiency \(\omega_{xjt}\) displays a secular downward trend, from -0.014 in 2003 to -0.066 in 2015 as shown in Figure \ref{fig:org667477a}. The falling matching efficiency suggests that the gap of quality between top workers and non-top workers narrowed over the period 2003-2015, noting that the quality of top workers is higher than that of non-top workers. Measures of the cross-firm dispersion of match efficiency across firms also declined before picking up in 2015.

We also tried the estimation of a more general form of the matching function, \(\ln y_{jt} = b_0 + \omega_{xjt} + b_1\cdot \ln x_{jt}\), where we do not force \(b_1=1\). Using the ordinary least squares (OLS) estimation on the equation, we find that the estimate of \(b_1\) is virtually 1 on the pooled sample data, and is close to 1 for most sectors when we estimate the equation by sector. This result suggests that the assumptions of the model appear to be supported by the data.
\subsection{Measures of firm-level productivity}
\label{sec:orga2e7032}
With the coefficient estimates of production function, we recover the total factor productivity (in natural logarithm), calculated as
\[\omega_{jt} = \widehat{\phi}_{jt} - \widehat{\beta}_0 - \widehat{\theta} \ln y_{jt} - \widehat{\alpha}_l \ln l_{jt} - \widehat{\alpha}_k \ln k_{jt},\]
where coefficient estimates are sector specific. In order to make comparison with productivity measured by statistical agency, we define two indexes of the measured productivity (in natural logarithm) as
\[z_{jt} = \omega_{jt} + \widehat{\theta} \ln y_{jt},\] and
\[z_{jt} = \omega_{jt} + \widehat{\theta} \omega_{xjt} + \widehat{\theta} \ln x_{jt}.\]
To be clear, we call \(z_{jt}\) the measured productivity, \(\omega_{jt}\) the total factor productivity in the context of our model, and \(\tilde{y}_{jt}=\widehat{\theta} \ln y_{jt}\) the contribution of top-worker quality to measured productivity. The estimate of output elasticity \(\widehat{\theta}\) is sector specific. Both indexes capture the contribution of worker quality to productivity. For firms within the same sector, the two measured indexes are parallel and differ only by a constant distance. The distance between the two indexes varies with sector, because the intercept term in the empirical matching function differs by sector.

The measured aggregate productivity is slightly different between the two indexes, and we use the measure based on the top-worker quality for productivity analysis. Thus, we must make it clear that although the subsequent analysis discusses the role of top workers in measured productivity growth, it is equivalent to the roles of match efficiency and non-top workers because the matching between top workers and non-top workers is positive and linear. We use the measured productivity based on top-worker quality because the match efficiency declined over the sample period and such a decline is due to the faster quality declines of top workers relative to non-top workers.
\section{Aggregate productivity and worker quality}
\label{sec:org8ebb5a7}
\subsection{Productivity slowdown and reallocation}
\label{sec:org7b71d90}
The measured aggregate productivity is defined as \(e^{z_t} = e^{\omega_t} \cdot e^{\tilde{y}_t}\). Each component is the geometric mean of firm-level values. In logarithm, it is the weighted sum of firm-level productivity, \(z_t = \omega_t + \tilde{y}_t\), where \(\omega_t = \sum_{j}^{n_{t}} s_{jt}\omega_{jt}\), \(\tilde{y}_t= \sum_{j}^{n_{t}} s_{jt} \widehat{\theta} \ln y_{jt}\), and \(n_t\) is the number of firms in year \(t\). Weight \(s_{jt}\) is firm \(j\)'s share of current-price output (value added) in the current-price aggregate output. We note again that \(\widehat{\theta}\) is sector specific.

\begin{figure}[htbp]
\centering
\includegraphics[angle=0,scale=0.30]{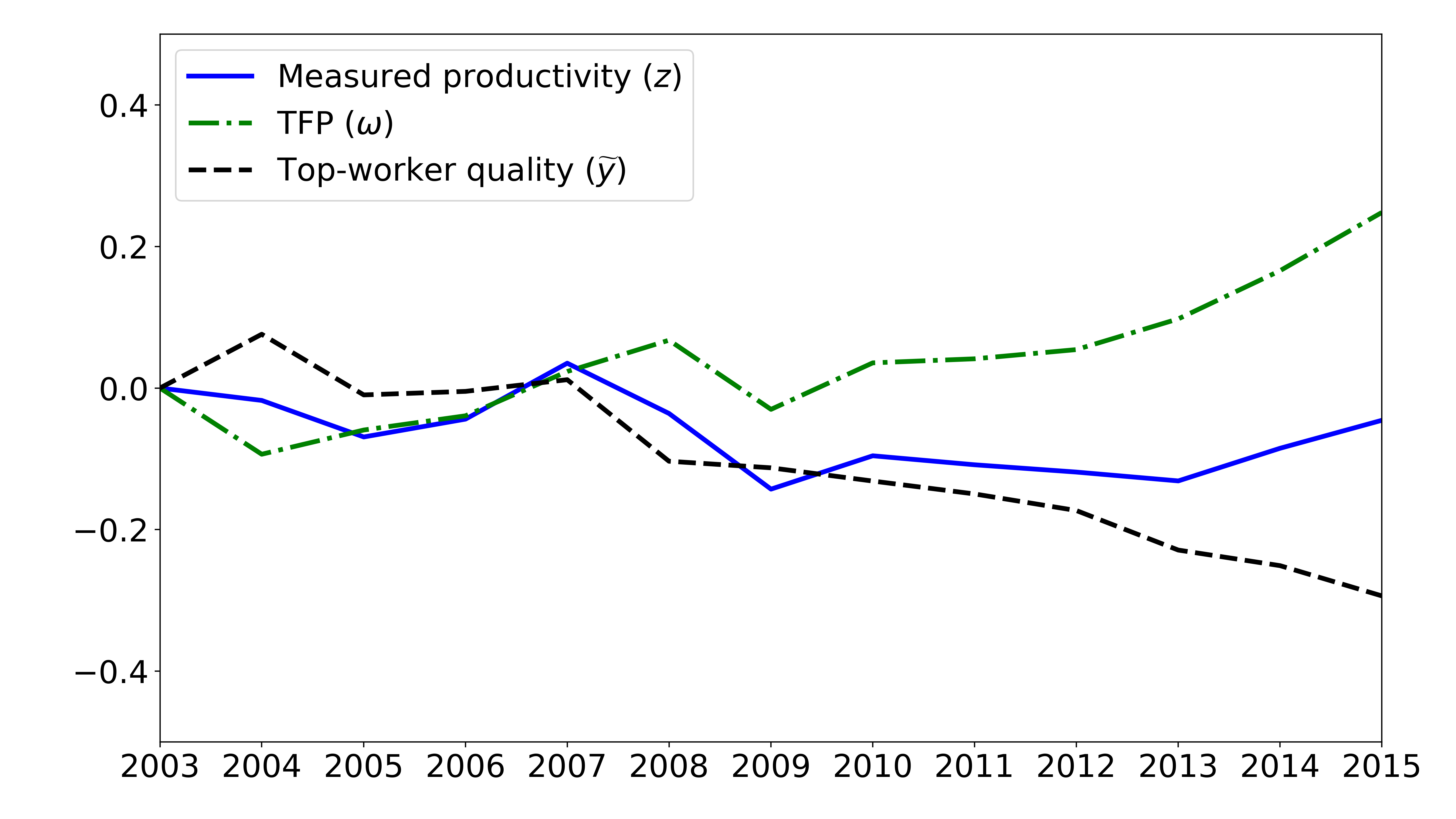}
\caption{\label{fig:org5904af9}Aggregate productivity measures}
\end{figure}

Figure \ref{fig:org5904af9} shows the logarithm of aggregate productivity measures, in which all three series in 2003 are normalized to zero. The measured aggregate productivity exhibits a secular downward trend over 2003-2015. It is 4.5 percent lower in 2015 than in 2003, which translates to an average of 0.38 percent of decline per year. In comparison, the multifactor productivity for the business sector, measured by Statistics Canada, is 4.0 percent lower in 2015 than in 2003, which translates to a 0.33 percent drop per year over the period 2003-2015. Our measured productivity overall displays a similar trend as the official measure that is based on sector-level data, despite some major differences in measurement.\footnote{There are two major differences. Our measured productivity excludes two sectors: Agriculture and Utilities. In our measure, labor input is the count of workers, while in measuring the multifactor productivity, Statistics Canada measures the labor input by taking into account educational composition of the work force.} In both measures, the aggregate productivity declined from 2003 to 2008, and its recovery after 2008 was so modest that its level is lower in 2015 than in 2003.

Furthermore, the decline of measured aggregate productivity is entirely accounted for by the contribution of top-worker quality. The estimated aggregate total factor productivity (Hicks-neutral technology) increases an average of 2.0 percent per year, while the component accounted for by the top-worker quality falls an average of 2.4 percent per year. The diverging trends of the two components lead to the declines of measured productivity. The Hicks-neutral technology growth sped up from 2013, with the last three years contributing the most to the technology progress over the sample period. Meanwhile, the component accounted for by the top-worker quality has been declining steadily since 2008.

Next, we find that the reallocation across firms plays a dominant role in the slowdown of aggregate productivity growth. Following \cite{olley1996}, we decompose the logarithm of aggregate measure \(z_t\) into two components
\[z_t = \overline{z}_t + \sum_{j=1}^{n_t} (s_{jt} - \overline{s}_t)(z_{jt} - \overline{z}_t),\]
where \(\overline{z}_t\) represents the unweighted mean value of the underlying measure over all firms, \(s_{jt}\) represents firm \(j\)'s output share in period \(t\), and \(\overline{s}_t\) is the firm-level average of shares. The second term on the right hand side of the equation is the covariance of output shares and the underlying measure (we ignore the term \(1/n_t\)). This covariance term reflects the contribution of reallocation to aggregate productivity. When the covariance falls over time, it means that levels of output produced by more productive firms fall relatively, suggesting that factors of production flow away from these firms and into firms that are less productive.

Table \ref{tab:orgf7e149c} summarizes the decomposition for all three aggregate measures in terms of average annual growth rates. Over the period 2003-2015, the unweighted mean productivity grew 0.61 percent per year. This increase was offset by the negative contribution of reallocation, captured by the covariance between output shares and productivity, which fell close to 1 percent per year. The resulting growth of the measured productivity was -0.38 percent per year. Thus, the slowdown of measured aggregate productivity is fully accounted for by the reallocation term.\footnote{We also decomposed the measured aggregate productivity at the two-digit NAICS level, we find that reallocation across sectors facilitated productivity growth. Without reallocation, the measured aggregate productivity would have grown twice slower. The positive contribution of cross-sector reallocation arises from total factor productivity, while cross-sector reallocation of top-worker quality is on average negative.} The covariance term fell in 2008 and stayed low. By 2015, it was 11.9 percent lower than in 2003. The reduced covariance suggests that firms that were more productive and had a better quality of top workers produced relatively less over time, a sign that firms with a higher measured productivity attracted fewer workers and invested less, relative to less productive firms.

\begin{table}[htbp]
\caption{\label{tab:orgf7e149c}Average annual growth rates (percent)}
\centering
\begin{tabular}{l|r|r|r}
\hline
\hline
 & 2003-2015 & 2003-2008 & 2008-2015\\
\hline
Aggregate Measured Productivity & -0.38 & -0.72 & -0.14\\
\(\ \text{    }\) Unweighted average & 0.61 & 0.65 & 0.58\\
\(\ \text{    }\) Covariance & -0.99 & -1.37 & -0.72\\
\hline
Aggregate TFP & 2.07 & 1.35 & 2.58\\
\(\ \text{    }\) Unweighted average & 1.58 & 1.47 & 1.66\\
\(\ \text{    }\) Covariance & 0.49 & -0.12 & 0.92\\
\hline
Aggregate contribution of top worker & -2.45 & -2.07 & -2.71\\
\(\ \text{    }\) Unweighted average & -0.97 & -0.82 & -1.07\\
\(\ \text{    }\)  Covariance & -1.48 & -1.24 & -1.64\\
\hline
\end{tabular}
\end{table}

The negative contribution of top-worker quality to measured productivity is reflected in both the unweighted average productivity and the reallocation terms. The decomposition of measured aggregate productivity can be further written as
\[z_t = \overline{\omega}_t + \sum_{j=1}^{n_t} (s_{jt} - \overline{s}_t)(\omega_{jt} - \overline{\omega}_t) + \overline{\tilde{y}}_t + \sum_{j=1}^{n_t} (s_{jt} - \overline{s}_t)(\tilde{y}_{jt} - \overline{\tilde{y}}_t),\]
where \(\overline{\tilde{y}}_t\) is the unweighted mean value of \(\tilde{y}_{jt}\). Applying this decomposition to our estimated series, we find that the unweighted mean of the contribution of top worker quality offsets most of the positive growth of the average Hicks-neutral technology, while the covariance with output shares fell for the contribution of top-worker quality but edged up for the Hicks-neutral technology, as shown in Table \ref{tab:orgf7e149c}. On net, the slowdown of measured productivity is driven by the contribution of top-worker quality. Reallocation played a bigger role than the unweighted mean in the declines of the contribution of top-worker quality.
\subsection{The negative contribution of top-worker quality}
\label{sec:org9abc0d2}
\begin{figure}[htbp]
\centering
\includegraphics[angle=0,scale=0.30]{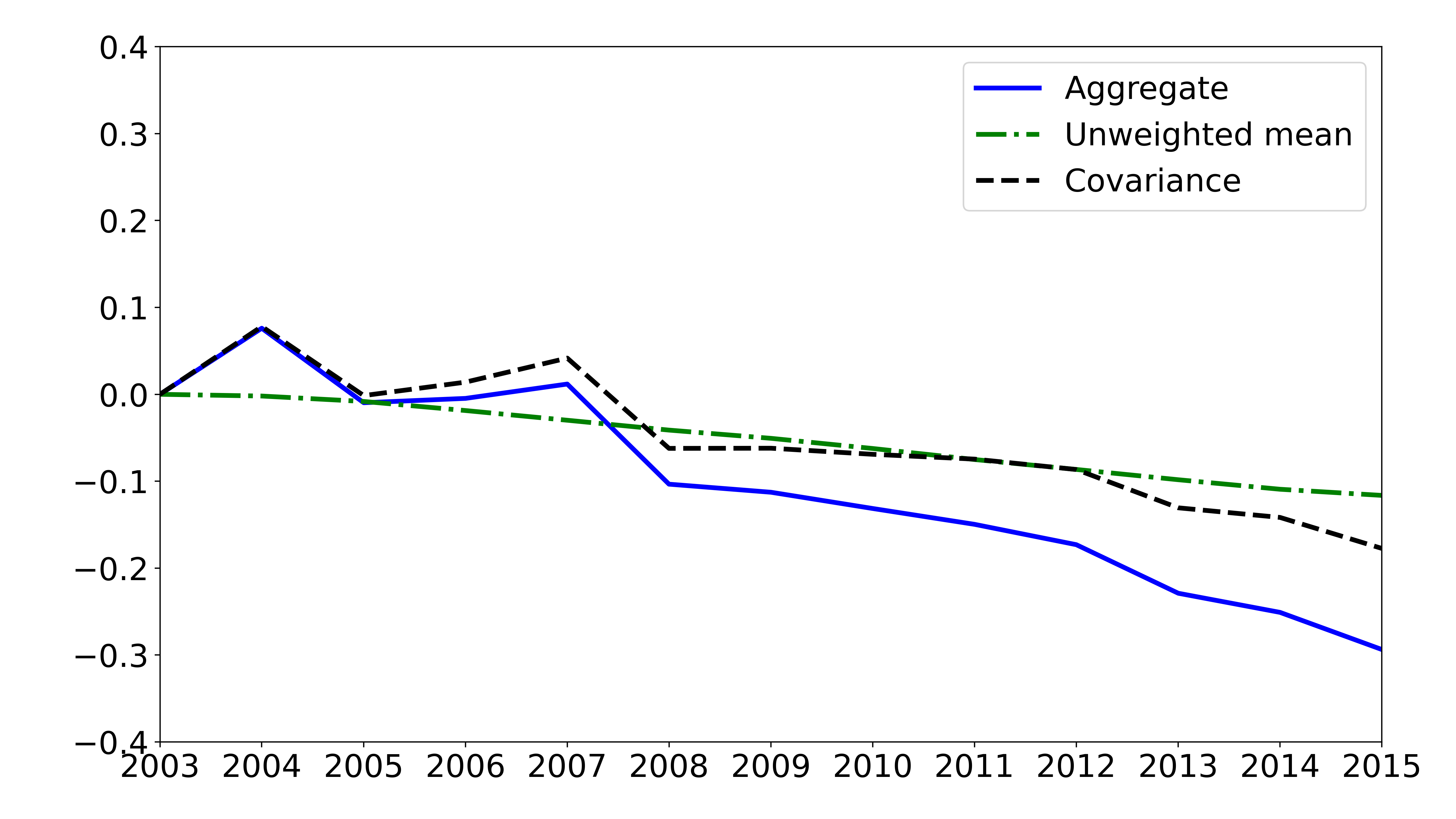}
\caption{\label{fig:org076d1b8}Decomposing the contribution of top-worker quality}
\end{figure}

The negative contribution of the top-worker quality to the growth of measured aggregate productivity is attributable to several related trends. First, both the unweighted average contribution of top-worker quality and the covariance between output shares and firm-level top-worker quality declined over the sample period, as is shown in Figure \ref{fig:org076d1b8}. Reallocation outweighs the firm-level average. About 60 percent of the contribution of top-worker quality to measured productivity is associated with the falling reallocation term, i.e. over time, firms with a higher top-worker quality produce less relatively.

Second, the negative contribution of the top-worker quality to the slowdown of measured productivity growth is associated with the falling aggregate top-worker quality since the output elasticity \(\widehat{\theta}\) does not vary over time. The quality of top workers dropped more than that of non-top workers. On average, the aggregate quality measure dropped 1.8 percent per year for non-top workers and 2.6 percent per year for top workers. The gap between these declines is captured by the match efficiency in the light of the matching function in our model. Though the unweighted average match efficiency declined over the period 2003-2015, the aggregate match efficiency (weighted sum) was increasing before 2008 and declining since then. About one third of the decline in top-worker quality is associated with the overall falling match efficiency. This means that the gap of quality between top workers and others narrowed since 2008.

Top-worker quality has two components, the person fixed effect and the age-sex effect. The person fixed effects play the most significant role in the fall of aggregate top-worker quality. On average, the aggregate fixed effect of top workers fell 2 percent per year, accounting for 78 percent of the decline of top-worker quality. This average contribution is concentrated in years since 2008, while the aggregate top-worker fixed effect was virtually flat during the period of 2003-2008. In 2015 alone, the aggregate top-worker fixed effect dropped 14.7 percent. Applying the Olley-Pakes decomposition to the aggregate top-worker fixed effect, we find that overall, the unweighted mean of top-worker fixed effect is the main source of the decline in aggregate top-worker fixed effect over the period 2003-2015, according to Table \ref{tab:orga69f8c8}. This contribution is uneven before and after 2008, though the decline of the average top-worker fixed effect was about the same in the two sub-periods. In the sub-period 2003-2008, the covariance between output shares and top-worker fixed effect rose, but worsened sharply after 2008. This again suggests that it is the deterioration of reallocation that is the main force driving the slowdown of measured productivity since 2008.

\begin{table}[htbp]
\caption{\label{tab:orga69f8c8}Aggregate top-worker quality, average annual growth rates (percent)}
\centering
\begin{tabular}{l|r|r|r}
\hline
Period & 2003-2015 & 2003-2008 & 2008-2015\\
\hline
Aggregate top-worker quality & -2.60 & 0.55 & -4.85\\
\(\ \text{    }\) Unweighted average & -1.56 & -1.26 & -1.77\\
\(\ \text{    }\) Covariance & -1.04 & 1.81 & -3.08\\
\hline
Aggregate top-worker fixed effect & -2.03 & 0.84 & -4.08\\
\(\ \text{    }\) Unweighted average & -1.43 & -1.38 & -1.47\\
\(\ \text{    }\) Covariance & -0.60 & 2.22 & -2.61\\
\hline
\end{tabular}
\end{table}

\begin{figure}[htbp]
\centering
\includegraphics[angle=0,scale=0.30]{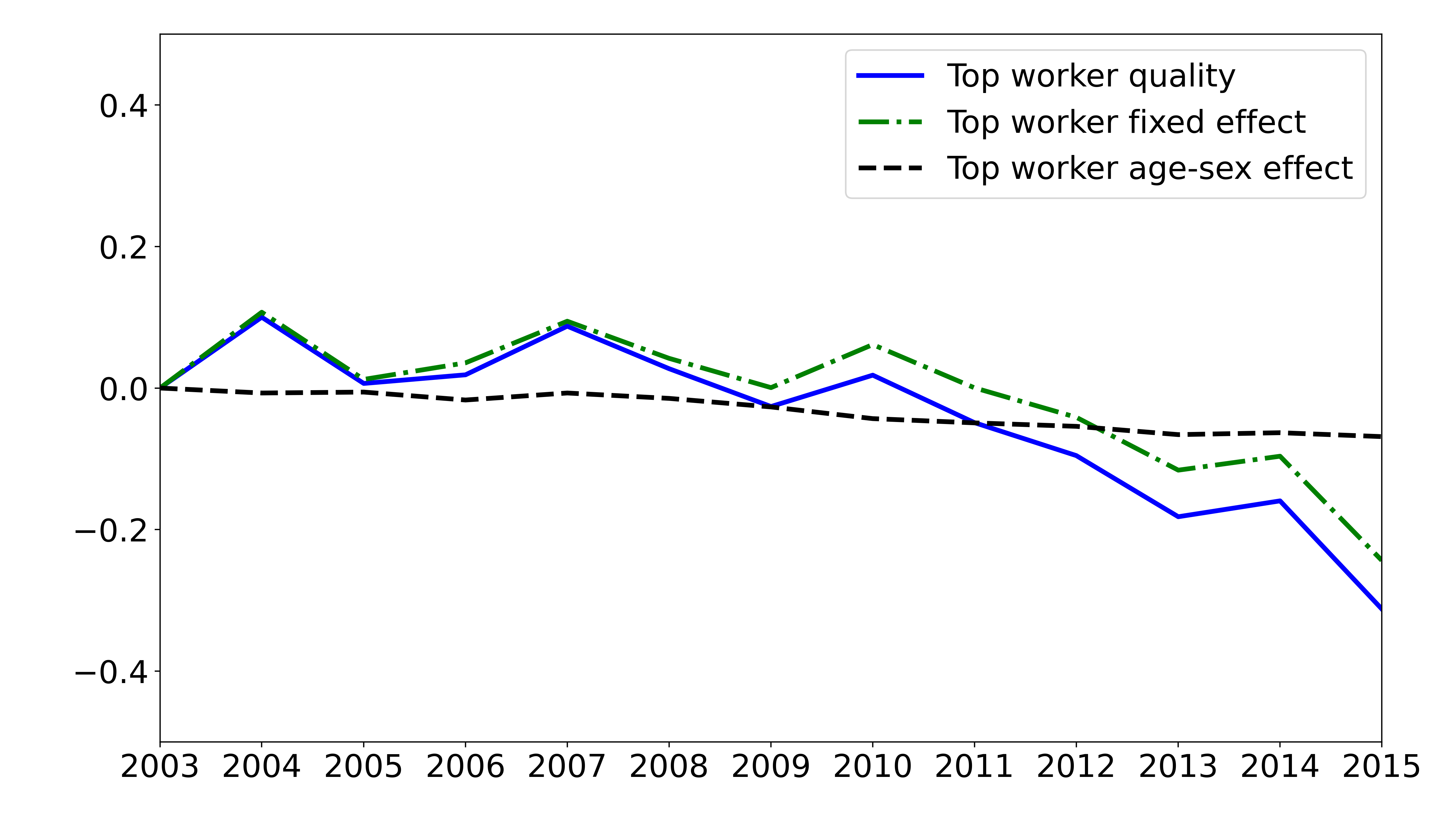}
\caption{\label{fig:orgfa3644f}Aggregate top-worker quality and age-sex effect}
\end{figure}

The falling aggregate top-worker quality also points to the effects related to worker age and sex. Worker quality measured from the estimated AKM earnings decomposition consists of the worker fixed effect and a component that we label as the age-sex effect. This age-sex effect refers to estimates of the interaction of age variables and sex in accounting for labor income variation. It observes a steady downward trend over the sample period, at an average rate of 0.57 percent per year (see Figure \ref{fig:orgfa3644f}). The aggregated age-sex effect for top workers was 6.9 percent lower in 2015 than in 2003. This drop accounts for 22 percent of the decline in the aggregated top-worker quality between the two years. The declining age-sex effect is a sign of the impact of population aging in Canada where baby boomers first reached retirement ages around 2010. The declining age-sex effect is consistent with the hump-shaped life-cycle profile of labor earnings found in micro-level data.

\subsection{Productivity dispersion and top-worker quality}
\label{sec:org9a59086}
\begin{figure}[htbp]
\centering
\includegraphics[angle=0,scale=0.30]{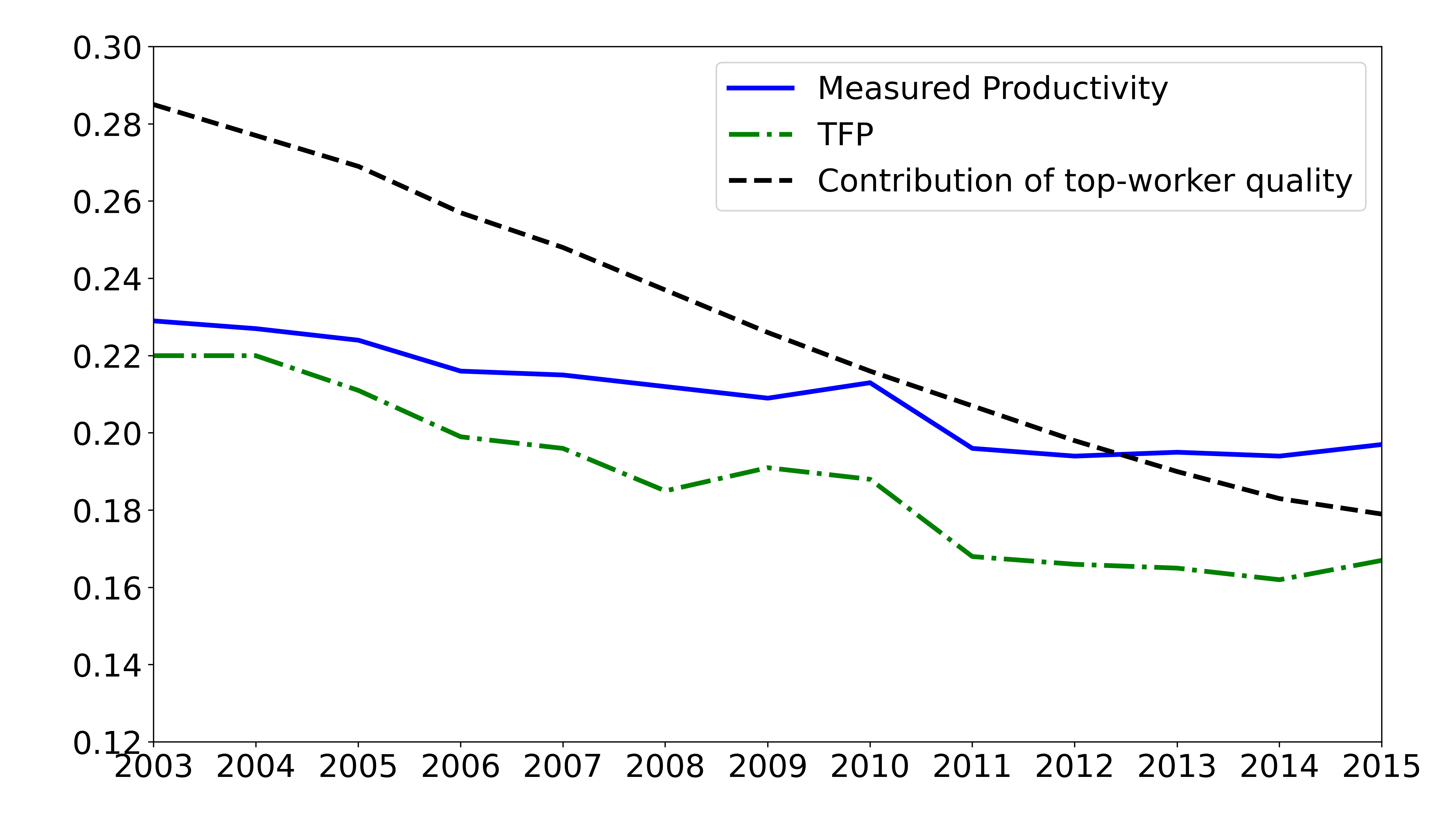}
\caption{\label{fig:orgbae763e}Variance of aggregate measures}
\end{figure}

We now examine to what extent the heterogeneity in the decline of top-worker quality contributed to the dynamics of productivity dispersion. From 2003 to 2015, the dispersion of measured productivity across firms has fallen. For example, the variance of the logarithm of all three measures declined, as shown in Figure \ref{fig:orgbae763e}. The covariance between total factor productivity and top-worker quality was negative and has been weakening over the sample period. The reduction of the variance of top-worker quality dominated that of measured productivity, contributing the most to the narrowing dispersion of measured productivity.

The falling dispersion of measured productivity is also found in both the interquartile range and the 90th-10th percentile difference. From 2003 to 2015, the 90th-10th percentile difference reduced by 12 percent, which was mainly driven by the faster productivity growth of firms at the bottom of productivity distribution. Measured productivity increased 15 percent from 2003 to 2015 among the 10th-percentile firms, but merely 2.5 percent among the 90th-percentile firms. This points to another pattern: that the slowdown of growth of the measured aggregate productivity appears to from firms at the top of the productivity distribution.
\subsection{Productivity under Cobb-Douglas production function}
\label{sec:org3fa7cd4}
The findings above are based on production function estimation in which we circumvented estimating the elasticity of substitution (\(\sigma\)). An alternative is to assume that the "output" function of the match takes the Cobb-Douglas form. For this alternative, we derived the optimal matching and obtained the measured productivity, as reported in Appendix \ref{sec:org374f69b}. We confirm that the patterns of trends of productivity and the role of worker quality in the Cobb-Douglas case are quantitatively similar to (and qualitatively the same as) the findings reported in the sections above.

There is one difference in terms of accounting for measured productivity slowdown. We show that, if the "output" function of the match is Cobb-Douglas, the optimal matching between top workers and non-top workers is a nonlinear function and depends on the Hicks-neutral technology. This is because the Hicks-neutral technology is also a measure of match efficiency, therefore the two are not distinguishable. As a result, in the measured productivity we may not cleanly separate the Hicks-neutral technology from the contribution of worker quality, and we explain this in more detail in Appendix \ref{sec:org374f69b}.
\section{Why did the top-worker quality decline?}
\label{sec:org39cd343}
The basic facts concerning the top workers at the aggregate level are: First, aggregate top-worker quality declined over the period 2003-2015 but mostly since 2008; Second, the person fixed effect of top workers in AKM estimation accounted for most of the declines of top-work quality. Reallocation of top-worker quality is at least as important as the unweighted average of top-worker quality in those declines. Why did the top-worker quality decline in particular from 2008 to 2015? Understanding causes of the decline requires us to examine further facts related to top talents in the labor market, which is beyond this paper. Nevertheless, we provide two possible reasons that can potentially explain the declines.

One explanation more plausible than others is that the emigration of top talents from Canada to the countries (mostly the United States) may have contributed to the stagnant top-worker quality observed in the data. \cite{macgee2025} find that the top 10 percent of the income distribution accounts for about two-thirds of measured labor productivity gap between Canada and the United States. According to the authors' calculations, the estimated net emigration flow of top earners from Canada (brain drain) to the United States dropped from 1960 to 2001 and started to increase in 2001, coinciding with rising and falling top income inequality. With the narrowing and widening gaps of measured labor productivity between Canada and the U.S. Based on standard neoclassic growth theory, the authors find that emigration of top earners may play a significant role in accounting for gaps in top income and in labor productivity.\footnote{Top earners in \cite{macgee2025} are defined by ranking all earners in the labor force, while top workers in this paper are workers with the highest pay in each firm. Nevertheless, top earners in the labor force and top workers in the firm population must overlap.} The widening Canada-U.S. gap of measured labor productivity growth in the 2000s corresponds to the slowdown of measured productivity growth in Canada.

Another plausible contributor to the declining top worker quality could be the taxation on top earners. Share of income tax paid by the top 10 percent earners in total income tax, relative to the share of their market income in aggregate income, has been rising until 2009. In 1992, the share of income tax paid by the top 10 percent earners was 46.9 percent of total income tax, while the share of market income of the top 10 percent earners was 35.7 percent of total market income, a difference of 11.2 percentage points. By 2009, the income tax share reached 54.8 percent and the market income share rose to 39.7 percent, a difference of 15.1 percentage points.\footnote{See Statistics Canada Table 11-10-0055, High income tax filers in Canada.} Starting from 2009, this relative share became roughly flat before picking up again in 2017. The increases of income tax paid relative to the income growth may make emigrating to the United States more attractive and dampen productivity of top workers. However, whether taxation on top workers is quantitatively important for stagnant top-worker quality needs a further investigation.

It could be also possible that increases in creative destruction by new entrepreneurs (entrants) outpaced innovation by top earners, hence lowering the top worker income, according to \cite{jones2018}. However, this explanation is inconsistent with another secular trend in Canada, that is, the firm entry rate has been declining in Canada as documented by \cite{cao2017cpp}.
\section{Conclusion}
\label{sec:org9fe16b3}
We incorporate worker quality and worker sorting into firms' production function and quantify the role of worker quality in firm productivity. If worker types display a constant elasticity of substitution in their contribution to production and if the distributions of worker types are Pareto, we show that there is a positive assortative matching between a firm's top worker and non-top workers. Furthermore, the matching function is linear in worker types with a closed form.

Taking into account the optimal matching between top workers and others, we estimate the production function for Canadian firms over the period 2003-2015, from which we obtain the firm-level total factor productivity (Hicks-neutral technology). We then obtain the measured total factor productivity which consists of the firm's total factor productivity and the term related to the quality of top workers. Aggregating these measures, we find that measured aggregate productivity declined slightly from 2003 to 2015, consistent with that estimated by the statistical agency. Such slowdown of productivity growth, according to our model, is entirely attributed to the downward trend of the aggregate top-worker quality, while the estimated Hicks-neutral technology has been rising over the sample period. We further show that reallocation played a dominant role in the falling measured productivity, largely due to the declines of top-worker quality over the sample period.

Two important issues are left for future research. First, in our model, the coefficient for the elasticity of substitution between worker types is not identified in the structural model because the matching function does not involve the elasticity of substitution. How to fully estimate the production function with optimal worker matching remains a question to study further. The second issue beyond this paper is that the causes of declining top-worker quality remains largely unknown. Existing research points to emigration of top talents, but there is a lack of direct evidence and structural study. Other causes such as taxation of top workers and the changing market structure of the labor market may also be plausible.

Our findings imply that, to boost productivity growth, policy should address the issue of top-worker quality. If innovation and ideas are mostly created by top workers, effective policy tools should be designed and employed to retain top talents and encourage their creation activities. Such policy can include appropriate taxation on top income.

\clearpage

\bibliographystyle{apalike}

\bibliography{matchQualityBib}

\clearpage

\appendix

\textbf{Appendices}
\section{Derivation of analytical form of solution}
\label{sec:org1484cd0}
\subsection{Conditions for positive assortative matching}
\label{sec:orge7d5513}
For convenience, we let \(q(x,y) = \alpha_x (e^{\omega_x} x)^{1-\sigma} + \alpha_y y^{1-\sigma}\). Guided by \cite{eeckhout2018}, we derive the condition for positive assortative matching (PAM). The first-order necessary conditions with respect to \(x\) and \(l\) are
\[f_x(\omega, x,T(x),l(x)) - w'(x) l(x) = 0, \ \  f_l(\omega, x,T(x),l(x)) - w(x) =0.\]

The second-order (sufficient) condition for optimal choices is that the Hessian matrix is negative definite,
\[\mathbf{H} = \begin{bmatrix} f_{xx} - w''(x) l(x) & f_{xl} - w'(x) \\ f_{lx} - w'(x) & f_{ll} \end{bmatrix}.\]

This requires that
\[f_{xx} - w''(x) l(x) < 0 \ \text{ and } \ \det (\mathbf{H})= f_{ll} \cdot [f_{xx} - w''(x) l(x)] - [f_{xl} - w'(x)]^2 > 0.\]

The two inequalities imply that \(f_{ll}<0\).

Taking into consideration the equilibrium outcome, differentiate \(f_l(\omega, x,T(x),l(x)) - w(x) =0\) with respect to type \(x\), we get
\[f_{xl} + f_{yl} T'(x) + f_{ll} l'(x) - w'(x) = 0.\] Also differentiate the optimality condition \(f_x(\omega, x,T(x),l(x)) - w'(x) l(x) = 0\) with respect to type \(x\), we get
\[f_{xx} + f_{xy} T'(x) + f_{xl}l'(x) - w''(x) l(x) - w'(x) l'(x) = 0.\]

The two second-order derivatives lead to \(f_{xl} - w'(x) = - f_{yl} T'(x) - f_{ll} l'(x)\) and \(f_{xx} - w''(x) l(x) = -f_{xy} T'(x) - f_{xl}l'(x) + w'(x) l'(x)\). Substituting these two equations in the determinant of \(\mathbf{H}\), and also using the optimality condition \(f_x(\omega, x,T(x),l(x)) - w'(x) l(x) = 0\), we get
\[-f_{ll} f_{xy} T'(x) - f_{ll} l'(x) [f_{xl} - w'(x)] - [f_{xl} - w'(x)]^2 > 0,\]
\[-f_{ll} f_{xy} T'(x) - [f_{xl} - w'(x)]\cdot [f_{ll} l'(x) + f_{xl} - w'(x)] > 0,\]
\[-f_{ll} f_{xy} T'(x) + [f_{xl} - w'(x)]\cdot f_{yl} T'(x) > 0,\]
\[-T'(x) [f_{ll} f_{xy} - f_{xl}f_{yl} + f_{yl} f_x/l(x)] > 0.\]

Positive assortative matching means that \(T'(x) > 0\). For this to hold, the necessary condition is
\[[f_{ll} f_{xy} - f_{xl}f_{yl} + f_{yl} f_x/l(x)] \le 0.\]

Noting \(f_{ll}<0\), divide the expression above by \(f_{ll}\), the necessary condition for PAM becomes
\[f_{xy} - \frac{f_{xl}f_{yl} - f_{yl} f_x/l(x)}{f_{ll}} \ge 0.\]

Applying the form of production function, the condition above is reduced to
\[e^{\omega} [q(x,y)]^{\frac{\theta}{1-\sigma}-2} e^{\omega_x(1-\sigma)}\alpha_x \alpha_y (xy)^{-\sigma} \theta (\sigma-1)  \ge 0.\]

Thus, PAM requires either \((\theta > 0, \sigma\ge 1)\) or \((\theta <0, \sigma \le 1)\).
\subsection{Model solution under Pareto distributions}
\label{sec:org4bf5ee2}

If matching is positively assortative, the first-order necessary conditions are given by
\[f_x - w'(x) l(x) = 0; \quad f_l - w(x) =0; \quad T'(x) = \frac{\mathcal{H}(x)}{l(x)},\]
where \(\mathcal{H}(x) = g(x) / h(T(x)) = C \cdot \frac{[T(x)]^{\lambda_y+1}}{x^{\lambda_x+1}}\) with \(C = \frac{\lambda_x \underline{x}^{\lambda_x}}{\lambda_y \underline{y}^{\lambda_y}}\).
Taking the total differentiation of \(f_l - w(x) = 0\) with respect to \(x\), it leads to
\[f_{lx} + f_{ly} \cdot T'(x) + f_{ll} l'(x) = f_x / l(x)\]
where we have applied the condition \(w'(x) = f_x / l(x)\) on the right-hand side. Given the form of production function, we notice that \(f_{lx} = \frac{\alpha_l f_x}{l(x)}\), \(f_{ly} = \frac{\alpha_l f_y}{l(x)}\), and \(f_{ll} = \frac{(\alpha_l-1) f_l}{l(x)}\). We then have
\[(\alpha_l-1) f_x + \alpha_l f_y T'(x) = (1-\alpha_l) f_l l'(x).\]

Applying to the equation above the expressions for marginal products of respectively \(x, y\) and \(l\), we obtain
\begin{equation}
\label{eq:TX}
(\alpha_l-1) \theta \alpha_x (e^{\omega_x})^{1-\sigma} x^{-\sigma} + \alpha_l \theta \alpha_y y^{-\sigma} T'(x) = (1-\alpha_l) \alpha_l q(x,y)\frac{l'(x)}{l(x)}.
\end{equation}

We now obtain \(\frac{l'(x)}{l(x)}\). Total differentiating the optimal condition with PAM, \(T'(x) l(x) = \mathcal{H}(x)\), with respect to \(x\), we obtain \(\frac{l'(x)}{l(x)} = \frac{\mathcal{H}'(x)}{\mathcal{H}(x)} - \frac{T''(x)}{T'(x)}\). Given the Pareto distributions of worker types, \(\frac{\mathcal{H}'(x)}{\mathcal{H}(x)} = (\lambda_y+1)\frac{T'(x)}{T(x)} - (\lambda_x+1)\frac{1}{x}\). Thus,
\[\frac{l'(x)}{l(x)} = (\lambda_y+1)\frac{T'(x)}{T(x)} - (\lambda_x+1)\frac{1}{x} - \frac{T''(x)}{T'(x)}.\]

Note, this expression for \(\frac{l'(x)}{l(x)}\) is the same as that when the optimal condition is \(T'(x) l(x) = -\mathcal{H}(x)\) (under NAM).

Thus, Equation (\ref{eq:TX}) becomes
\[(\alpha_l-1) \theta \alpha_x (e^{\omega_x})^{1-\sigma} x^{-\sigma} + \alpha_l \theta \alpha_y y^{-\sigma} T'(x) = (1-\alpha_l) \alpha_l q(x,y)\left[(\lambda_y+1)\frac{T'(x)}{T(x)} - (\lambda_x+1)\frac{1}{x} - \frac{T''(x)}{T'(x)}\right].\]
It can be written as
\[\frac{\theta}{\alpha_l} + \frac{\theta}{1-\alpha_l}\cdot \frac{T'(x)}{T(x)}x - \left[1+\frac{\alpha_y}{\alpha_x (e^{\omega_x})^{1-\sigma}}\left(\frac{T(x)}{x}\right)^{1-\sigma}\right]\cdot \left[\left(\frac{\theta}{1-\alpha_l}-\lambda_y - 1\right) \frac{T'(x)}{T(x)}x + \lambda_x+1+\frac{T''(x)}{T'(x)}x\right]=0.\]
This is a second-order differential equation. Guess \(y = T(x) = A x^B\) with unknown parameters \(A\) and \(B\). Using the guessed functional form, the equation above becomes
\[\frac{\theta}{\alpha_l} + \frac{\theta}{1-\alpha_l}\cdot B - \left[1+\frac{\alpha_y}{\alpha_x (e^{\omega_x})^{1-\sigma}}\left(Ax^{B-1}\right)^{1-\sigma}\right]\cdot \left[\left(\frac{\theta}{1-\alpha_l}-\lambda_y - 1\right) B + \lambda_x+1+B-1\right]=0.\]

With the CES matching function, \(\sigma\neq 1\), it must be that \(B=1\) in order for the equation to hold. We can then solve for \(A\) as
\[A = \left[\frac{\alpha_x (\theta/\alpha_1 + \lambda_y - \lambda_x)}{\alpha_y(\theta/(1-\alpha_l) - \lambda_y + \lambda_x)}\right]^{\frac{1}{1-\sigma}} e^{\omega_x}.\]

The condition for positive assortative matching is \(A>0\), which requires that, for any \(\sigma\neq 1\):
\[ \theta > \max \left\{\alpha_l(\lambda_x-\lambda_y), -(1-\alpha_l)(\lambda_x-\lambda_y)\right\}\]
or
\[ \theta < \min \left\{\alpha_l(\lambda_x-\lambda_y), -(1-\alpha_l)(\lambda_x-\lambda_y)\right\}.\]

To ease notation, we write \(A = \Psi^{\frac{1}{1-\sigma}} \cdot e^{\omega_x}\). The equilibrium match function is then
\[y = T(x) = \Psi^{\frac{1}{1-\sigma}} e^{\omega_x} x.\]

Using the optimal condition for the matching market clearing, we solve for the optimal demand for labor, as
\[l(x) = \left(\Psi^{\frac{1}{1-\sigma}} \cdot e^{\omega_x}\right)^{\lambda_y} C x^{\lambda_y - \lambda_x}.\]

Using \(f_l - w(x) = 0\), we find that the equilibrium wage rate for type \(x\) is
\[w(x) = \Lambda \cdot e^{\omega} \cdot e^{\omega_x(\theta - \lambda_y(1-\alpha_1))} \cdot x^{\theta + (1-\alpha_l)(\lambda_x-\lambda_y)},\]
where \(\Lambda = \alpha_l (\alpha_x+\Psi \alpha_y)^{\frac{\theta}{1-\sigma}} \cdot \Psi^{\frac{\lambda_y(\alpha_l-1)}{1-\sigma}} \cdot C^{\alpha_l-1}\).

Taking into account the match in equilibrium, the production function becomes
\[f(x,T(x),l) = (\alpha_x + \alpha_y \Psi)^{\frac{\theta}{1-\sigma}} e^{\omega} (e^{\omega_x}x)^{\theta} [l(x)]^{\alpha_l}.\]

This production function does not identify output elasticity because both Hicks-neutral productivity \(\omega\) and the match efficiency \(\omega_x\) affect the firm's optimal choice on non-top worker type \(x\) and the work force, and both efficiency indexes are unobserved. Alternatively, the production function can be transformed to the following form
\[f\left(T^{-1}(y),y,l\right) =\left(\alpha_x \Psi^{-1}+ \alpha_y\right)^{\frac{\theta}{1-\sigma}} e^{\omega} y^{\theta} \left[l\left(T^{-1}(y)\right)\right]^{\alpha_l}.\]
We will estimate this production function and recover productivity \(\omega\) to analyze productivity dynamics.
\section{Data sample and descriptive statistics}
\label{sec:orge955ae1}
\subsection{Data source}
\label{sec:org4fa46d1}
We use the Canadian Employer Employee Dynamics Database (CEEDD) 2017 vintage, a data infrastructure regarding the Canadian population of businesses and persons from 2001 to 2015. The CEEDD consists in a set of data files that can be linked. We use the following data files:
\begin{itemize}
\item T1 personal master files, the individual tax files providing information on incomes, taxes, sex, birth year, and province;
\item T4-ROE files, the matched employer-employee data providing matched income, employer ID, employee ID, and NAICS;
\item Business ownership files, providing information on the share of ownership;
\item NALMF, a longitudinal data on enterprise income and financial statements.
\end{itemize}
\subsection{Data sample}
\label{sec:org10d2284}

Data are available from 2001 to 2015. We use data sample over the period 2003-2015 largely due to the cost of estimation and the limited computational power on the facility used for the project. In implementing the AKM estimation, we drop observations by imposing several restrictions.

First, drop a worker/year observation if the worker is younger than 20 or older than 64.

Second, T4-ROE records payments from an employer to a person including employment income. Income in many matches is small, and is often a one-time payment, which may not be based on a recurring employment relationship. We drop the matched income if it is below the provincial minimum hourly wage times 40 times 13.

Third, for workers with multiple matches in the same year, we keep the two matches with the highest incomes; Further, we drop the match with the lower income if this income is smaller than two-thirds of the higher income. The latter restriction helps mitigate the problem when a person moves in late months (see November) to a new job with a higher annual pay.

Fourth, we exclude the matches where employees are also an owner of the same business. We find that, in T4-ROE files, about 40 percent of firm/year observations over 2001-2008 saw that at least one employee is an owner in the same firm, and this share rose to 44 percent in 2008-2015. Further, an owner is also the top-paid employee in 32 percent of firm/year observations over 2001-2008, and this share rose to 36 percent over 2008-2015.

Finally, in the estimation, we include businesses in the public sector (two-digit NAICS 91), but drop all matches in this sector in subsequent analysis.

Frequencies of the final data sample are show in Table \ref{tab:org852040a}, employees of a firm who are owners of the same firm are dropped. More than 99 percent of firm-person-year observations in the full data set are kept in the largest connected set. About 30 thousand firms are not connected hence dropped, and these are tiny firms. The average firm size (number of matches in a given year) is about 16.6 while the median firm size is 3 workers. The mean person age is 41.5, and male workers account for 51 percent of the person-year observations. About 5 percent of person-years observe two matches, these are workers who hold two jobs or switch job.

\begin{table}[htbp]
\caption{\label{tab:org852040a}Frequencies (in millions), CEEDD 2003-1015}
\centering
\begin{tabular}{l|r|r|r|r|r}
\hline
\hline
Data set & Observations & Person-year & Unique persons & Firm-year & Unique firms\\
\hline
Full & 170.6 & 162.5 & 20.9 & 10.5 & 1.91\\
Connected & 169.0 & 161.0 & 19.5 & 10.2 & 1.62\\
\hline
\end{tabular}
\end{table}
\subsection{Job movements and income dynamics}
\label{sec:org0d24539}

Table \ref{tab:org8019987} shows the summary statistics of job moves and income. Overall, the average number of job moves is about 1.62 over the period 2003-2015. Close to six million workers did not change their jobs in the same period, accounting for 30 percent of unique workers. The median real earnings in logarithm increased 2.2 percent per year, while individuals experienced a median 2.0 percent increase in real earnings. The earnings growth is noisy as the observed earnings are job specific and not annualized, a person's observed earnings may observe a large change simply because the duration of two jobs is different.

\begin{table}[htbp]
\caption{\label{tab:org8019987}Job moves and income, CEEDD 2003-1015}
\centering
\begin{tabular}{l|l}
\hline
\hline
\# job moves & 31.68m\\
\# workers with no job move & 5.80m\\
Mean log earnings & 10.215\\
Variance of log earnings & 0.741\\
Log earnings 90th-10th difference & 2.28\\
Median change in earnings & 2.21\%\\
Mean change in earnings & 4.62\%\\
Median change in earnings on the same job & 2.0\%\\
Mean change in earnings on the same job & 3.16\%\\
\hline
\end{tabular}
\end{table}
\subsection{Top-paid workers}
\label{sec:org2fb0c6a}

We allow each employer (enterprise) to have at most two top-paid employees. In any given year, if the second highest matched income in an employer is more than 99.5 percent of the highest matched income, the employee with the second highest matched income is also counted as a top-paid worker. According to this definition of top-paid workers, less than 0.5 percent of firm-year observations have more than two top-paid workers. In the connected data set, 97.93 percent firm-year observations saw only one top-paid workers.

Over the period of 2003-2015, the average of logarithm of real matched earnings among top paid workers is 10.24. Top workers in larger firms earn more than those working in smaller firms. Top-paid workers in firms with 500 employees or more on average earn four times more than top-paid workers in firms with fewer than ten employees. The average of ratios of real matched earnings between the top-paid worker and the worker(s) paid the second highest is 1.7, while the median ratio is 1.3. This suggests that on average the top-paid worker earns 70 percent more than the worker paid the second highest. We also find that about 2 percent firm-year observations saw an earnings gap of 1 percent or smaller between the top worker and the worker paid the second highest.

\begin{table}[htbp]
\caption{\label{tab:org596344c}Top-paid workers, CEEDD 2003-1015}
\centering
\begin{tabular}{l|r}
\hline
\hline
Share of firm/year with 1 top-paid worker & 98\%\\
Mean Ratio of Top/Second Earnings & 1.7\\
Median Ratio of Top/Second Earnings & 1.3\\
Mean Ratio of Top/Non-Top Earnings & 2.6\\
Median Ratio of Top/Non-Top Earnings & 2.1\\
Share of firm/year with (Top/Second Earnings Ratio <1.01) & 2\%\\
\hline
\end{tabular}
\end{table}

\begin{figure}[htbp]
\centering
\includegraphics[angle=0,scale=0.30]{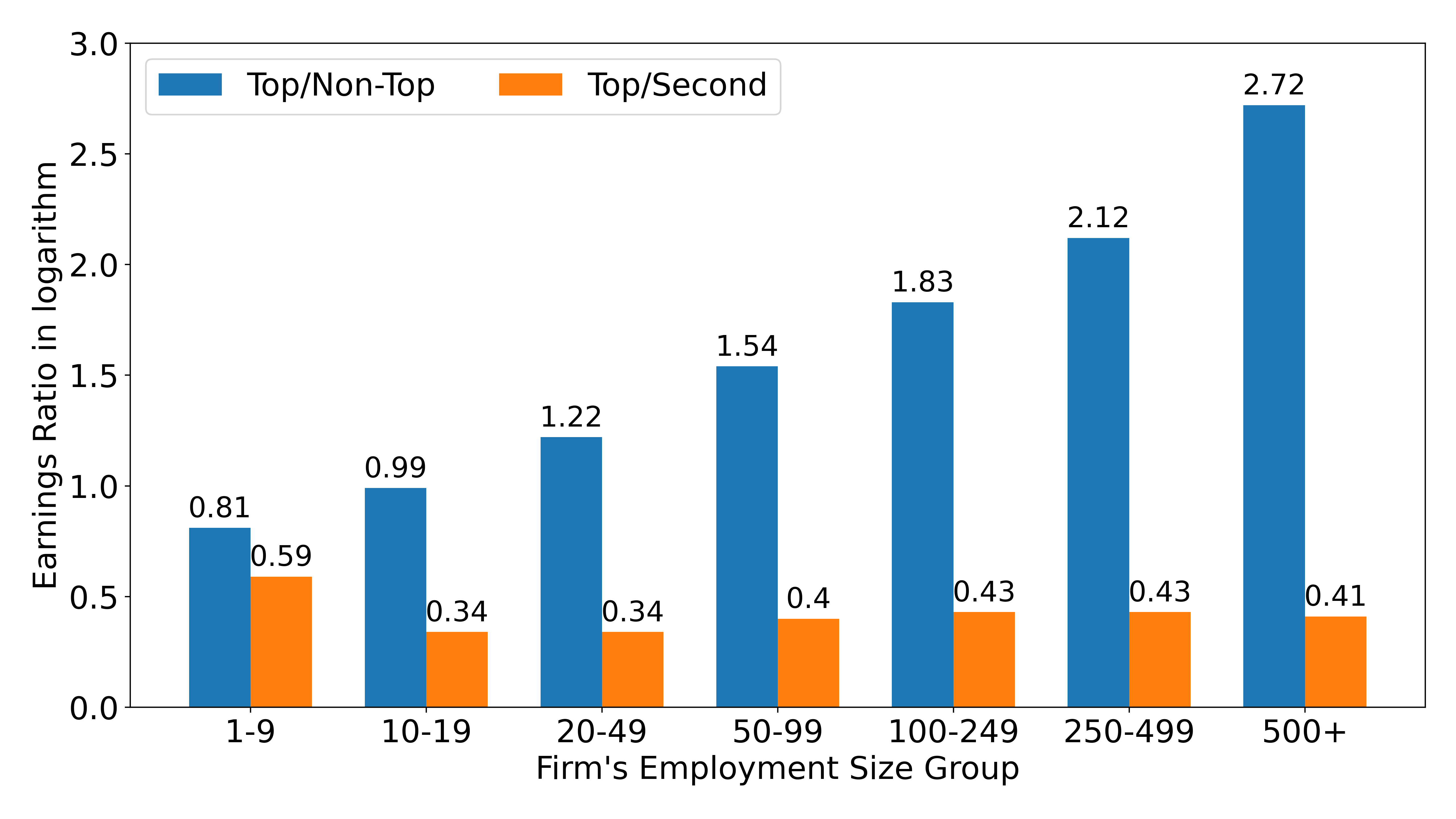}
\caption{\label{fig:orgcfa79eb}Earnings Ratio in Logarithm}
\end{figure}

The mean ratio of matched earnings between top paid workers and non-top workers differ across the firms' employment size, as shown in Figure \ref{fig:orgcfa79eb}. The ratio of earnings of top paid workers relative to the earnings of other workers in the same employer is about 2.25 among firms with fewer than 10 employees, in contrast, this ratio is 15 among firms with 500 employees or more. The earnings ratio between the top and second top paid workers displays small differences across firms' employment sizes.

\section{AKM estimation}
\label{sec:org1795b7d}
\subsection{Estimation}
\label{sec:orge90403f}
The two-sided fixed effects obtained through estimating the equation below:
\[\ln w_{ijt} = \beta_0 + \alpha_i + X_{it} \pmb{\beta} + \psi_{j(i,t)} + \mathrm{Yr}_t+ \varepsilon_{ijt},\]
where \(w_{ijt}\) is the annual matched earnings (in 2002 prices) of person \(i\) paid by firm \(j\) in year \(t\). \(\alpha_i\) is the person fixed effect and \(\psi_{j(i,t)}\) is the firm fixed effect. \(X_{it}\) includes a person's age squared and cubic, interacted with the dummy variable for male workers. \(Yr_t\) is a vector of two-year dummy variables, e.g., \(\mathrm{Yr}_{2004}=1\) if for matches in year 2003 and 2004.

The AKM estimation is implemented in Stata using the code by \cite{correia2017:HDFE}. Coefficient estimates are reported in Table \ref{tab:org326af23}. The R-squared is 0.756 and the adjusted R-squared is 0.721, which are in line with similar estimation that also uses annual matched income data such as \cite{song2019}.

\begin{table}[htbp]
\caption{\label{tab:org326af23}AKM Estimation}
\centering
\begin{tabular}{l|rrrr}
\hline
\hline
 & Coefficient & Std. Err. & {[}95\% Conf. Interval] & \\
\hline
age*Male & -0.098 & 0.0013 & -0.101 & -0.096\\
ageSqr & -2.370 & 0.0013 & -2.372 & -2.367\\
ageSqr*Male & -0.579 & 0.0018 & -0.583 & -0.576\\
ageCubic & 1.36 & 0.0032 & 1.354 & 1.366\\
ageCubic*Male & 0.589 & 0.0045 & 0.580 & 0.598\\
\(\text{Yr}_{2004}\) & -0.374 & 0.0003 & -0.375 & -0.374\\
\(\text{Yr}_{2006}\) & -0.337 & 0.0003 & -0.337 & -0.336\\
\(\text{Yr}_{2008}\) & -0.265 & 0.0002 & -0.266 & -0.265\\
\(\text{Yr}_{2010}\) & -0.203 & 0.0002 & -0.203 & -0.203\\
\(\text{Yr}_{2012}\) & -0.144 & 0.0002 & -0.145 & -0.144\\
\(\text{Yr}_{2014}\) & -0.068 & 0.0002 & -0.069 & -0.068\\
Constant & 10.65 & 0.0002 & 10.65 & 10.65\\
\hline
\end{tabular}
\end{table}
\subsection{Variance decomposition}
\label{sec:org019f3a3}
In documenting variance decomposition and correlations of fixed effects, we drop matches in NAICS 91 (public administration) from calculations. We first tabulate the variance decomposition using fixed effects at the match level, as in Table \ref{tab:orgba493e1}. It suggests that slightly more than 75 percent of variance of the logarithm of annual matched earnings is explained by the fixed effects and effects of age and sex. the worker fixed effect is by far the foremost important factor, accounting for more than 47 percent of the variance of matched earnings.

\begin{table}[htbp]
\caption{\label{tab:orgba493e1}Decomposition of dispersion in matched earnings}
\centering
\begin{tabular}{l|rrrrrr}
\hline
\hline
Firm Size & All & 1-9 & 10-19 & 20-99 & 100-499 & 500+\\
\hline
\# (person, year)'s & 152.1m & 22.8m & 12.2m & 29.8m & 23.0m & 64.4m\\
Var(ln w) & 0.748 & 0.610 & 0.670 & 0.709 & 0.708 & 0.713\\
Var(firmFE) & 0.092 & 0.132 & 0.067 & 0.059 & 0.058 & 0.068\\
Var(workerFE) & 0.355 & 0.346 & 0.323 & 0.340 & 0.341 & 0.359\\
Var(Xb) & 0.047 & 0.051 & 0.051 & 0.051 & 0.046 & 0.042\\
Var(residual) & 0.185 & 0.175 & 0.196 & 0.197 & 0.196 & 0.177\\
Cov(workerFE, firmFE) & 0.035 & -0.020 & 0.029 & 0.035 & 0.035 & 0.028\\
Cov(workerFE, Xb) & -0.007 & -0.018 & -0.011 & -0.008 & -0.009 & -0.004\\
Cov(firmFE, Xb) & 0.008 & 0.004 & 0.006 & 0.007 & 0.007 & 0.007\\
Var(workerFEXb) & 0.387 & 0.361 & 0.353 & 0.376 & 0.369 & 0.392\\
Cov(firmFE,workerFEXb) & 0.043 & -0.016 & 0.035 & 0.042 & 0.042 & 0.035\\
\hline
\end{tabular}
\end{table}

The correlation between firm fixed effects and worker fixed effects is weak, with a correlation efficient of 0.19, as also found in AKM estimation using other data. This correlation is negative for very small firms, suggesting that the pattern of assortative matching could be different between small and large firms.

Next, we split the work force of each employer into top workers and non-top workers, and obtain their fixed effects respectively. In doing so, the number of person-year observations drops to 141 million. A firm-year is dropped if the firm has only one workers in that year. Table \ref{tab:org354f9b5} suggests that the correlation is stronger between firm fixed effects and the top-paid worker fixed effects, than between firm and non-top workers.

\begin{table}[htbp]
\caption{\label{tab:org354f9b5}Dispersion and correlation of components of matched earnings}
\centering
\begin{tabular}{l|rrrrrr}
\hline
Firm Size & All & 1-9 & 10-19 & 20-99 & 100-499 & 500+\\
\hline
\# (person, year)'s & 141.0m & 15,7m & 10.6m & 28.2m & 22.6m & 64.0m\\
sd(firmFE) & 0.289 & 0.319 & 0.256 & 0.241 & 0.240 & 0.260\\
sd(topFE) & 1.308 & 0.589 & 0.603 & 0.706 & 0.770 & 1.172\\
sd(topFExb) & 1.249 & 0.579 & 0.574 & 0.661 & 0.724 & 1.145\\
sd(otherFE) & 0.581 & 0.522 & 0.530 & 0.555 & 0.571 & 0.596\\
sd(otherXb) & 0.217 & 0.235 & 0.228 & 0.227 & 0.214 & 0.205\\
sd(otherFExb) & 0.610 & 0.539 & 0.555 & 0.586 & 0.596 & 0.624\\
corr(firmFE, topFE) & 0.325 & 0.075 & 0.266 & 0.284 & 0.146 & -0.060\\
corr(firmFE, topFExb) & 0.316 & 0.091 & 0.279 & 0.286 & 0.137 & -0.072\\
corr(firmFE, otherFE) & 0.245 & 0.024 & 0.216 & 0.262 & 0.259 & 0.179\\
corr(firmFE, otherFExb) & 0.283 & 0.053 & 0.249 & 0.303 & 0.294 & 0.214\\
corr(topFE, otherFE) & 0.220 & 0.428 & 0.338 & 0.265 & 0.173 & 0.057\\
corr(topFExb, otherFExb) & 0.225 & 0.460 & 0.358 & 0.278 & 0.175 & 0.036\\
\hline
\end{tabular}
\end{table}
\subsection{Measuring worker quality}
\label{sec:org5db2095}
The top worker quality is measured as the top-paid worker person effect (fixed effects and those related to age and sex), specifically, the term \(\alpha_i + X_{it} \pmb{\beta}\) in the AKM estimation equation where \(i\) is the top-paid worker in firm \(j\) and year \(t\). The top worker quality can change over time due to top worker turnover. The non-top worker quality is measured as the average of fixed effects and those related to age and sex of non-top workers in the same firm/year. The non-top worker quality changes over time if the composition of the work force changes. In case a firm in the same year has two top-paid workers, we use as the top worker quality the average fixed effects and age-sex effects of the two top paid workers.

The correlation coefficient is 0.13 between the firm fixed effects and the top-worker person effects, according to Table \ref{tab:org5cd48b0}, while the correlation between the firm fixed effects and the non-top worker person effects is close to zero (0.07). However, the correlation of person effects between top and non-top workers is strong, with a coefficient 0.64. Not tabulated, we find that the Spearman rank correlation is about 0.35 between firm fixed effects and top-worker person effects, and 0.3 between firm fixed effects and non-top worker person effects.

\begin{table}[htbp]
\caption{\label{tab:org5cd48b0}Dispersion and correlation of components at the firm level}
\centering
\begin{tabular}{l|rrrrrr}
\hline
Firm Size & All & 1-9 & 10-19 & 20-99 & 100-499 & 500+\\
\hline
\# of firm/year's & 6.66m & 4.83m & 0.91m & 0.76m & 0.12m & 0.03m\\
Var(firmFE) & 0.116 & 0.129 & 0.068 & 0.060 & 0.058 & 0.061\\
Var(topFExb) & 0.479 & 0.353 & 0.324 & 0.414 & 0.511 & 0.793\\
Var(otherFExb) & 0.215 & 0.230 & 0.125 & 0.112 & 0.095 & 0.080\\
Cov(firmFE,topFExb) & 0.030 & -0.005 & 0.040 & 0.046 & 0.027 & 0.000\\
Cov(firmFE,otherFExb) & 0.012 & -0.010 & 0.037 & 0.044 & 0.043 & 0.030\\
Cov(topFExb, otherFExb) & 0.205 & 0.174 & 0.134 & 0.134 & 0.108 & 0.072\\
\hline
\end{tabular}
\end{table}
\section{Assumption on the distribution of worker quality}
\label{sec:org8c38dd7}
The analytical matching function is obtained under the assumption that both the top worker quality and the non-top worker quality have a Pareto distribution. This assumption is reasonable as it is supported by data. For random variable \(Y\), the probability \(P(Y>y) = (\underline{y}/y)^{\lambda_y}\). We can estimate the Pareto coefficient by the ordinary least squares (OLS) on the log-rank equation as below
\[\ln (\text{Rank} - 0.5) = \text{Constant} - \lambda_y \ln y + \varepsilon,\]
as suggested in \cite{gabaix2009}.

Applying the estimation to the top-worker quality, we use the pooled data for all years and add the year dummy to the equation. We obtain an estimate \(\widehat{\lambda}_y = 1.48\) with a standard error that is smaller than 0.0005 and the R-square is 0.927. The R-square is considered small, and is due to the top workers at the bottom. Visually, the log-rank of log-quality for top worker appears fairly flat at the bottom of worker quality distribution, ranks fall only slightly as worker quality rises. After passing a threshold value of the log-quality, the top worker quality observes a straight rank-quality relationship. For example, we obtain \(\widehat{\lambda}_y = 1.80\) and \(R^2=0.991\) when estimating the equation above for 2005 and dropping top workers with \(\ln y < -0.2\). The large \(R^2\) value suggests a power law of the top-worker quality. The threshold value -0.2 is at about the 15th percentile in 2005 (the median of \(\ln y\) is 0.24). A very similar result is obtained for the year 2010.

For non-top workers, the pooled estimation over all years leads to \(\widehat{\lambda}_x = 2.06\) (with standard error .001), but the R-square is 0.829. This suggests that the power law appears a stronger assumption for non-top workers than for top workers. The small R-square is also due to the bottom non-top workers. For example, when using non-top workers with \(\ln x >-0.4\) in the estimation, the results are \(\widehat{\lambda}_x = 3.40\) (stand error .001) and an R-square value 0.983, which is a reasonably good fit. But the threshold value -0.4 is large, fairly close to the median value of \(\ln x\). Overall, we consider that the Pareto distribution is a reasonable approximation for the distributions of worker quality.
\section{More on aggregate top-worker quality}
\label{sec:orgb144c34}
\begin{figure}[htbp]
\centering
\includegraphics[angle=0,scale=0.30]{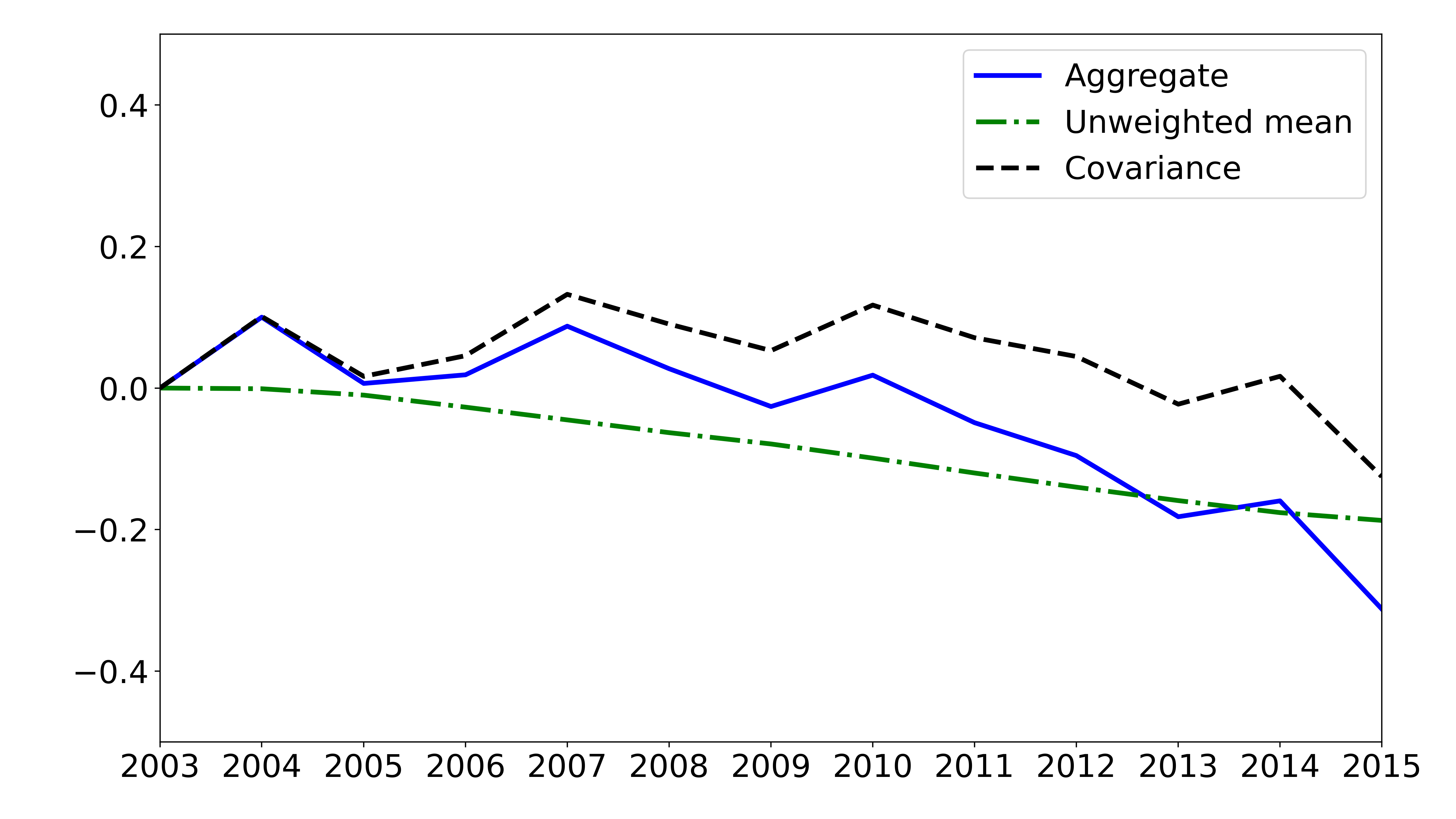}
\caption{\label{fig:org4b0492a}Decomposition of aggregate top-worker quality}
\end{figure}

In the main text, we decompose the contribution of top-worker quality, \(\tilde{y}_t\), to measured aggregate productivity. The aggregate top-worker quality itself, \(\sum_{i}^{n_t} s_{it} \ln y_{it}\), exhibits a fairly different pattern of trends, as shown in Figure \ref{fig:org4b0492a}. Its decline mainly comes from the falling quality of top workers at the individual firms (the unweighted mean), especially before 2015. The covariance term increases slightly over the years before 2015, suggesting that the output share and top-worker quality are positively correlated, firms with a higher quality level of top-worker tend to produce more. The covariance term over time captures the change in aggregate quality of top workers through reallocation of labor and capital from firms with low quality to firms with high quality level of top workers. This covariance nevertheless becomes smaller after the great recession of 2008, suggesting the slowdown of reallocation. On average, 40 percent of declines in aggregate quality of top workers is accounted for by the slow down in reallocation over 2003-2015. However, the contribution of reallocation to changes in aggregate top-worker quality is positive if we exclude the last year 2015. For the years before 2015, in the absence of reallocation, top worker quality would have declined slightly more.

\begin{figure}[htbp]
\centering
\includegraphics[angle=0,scale=0.30]{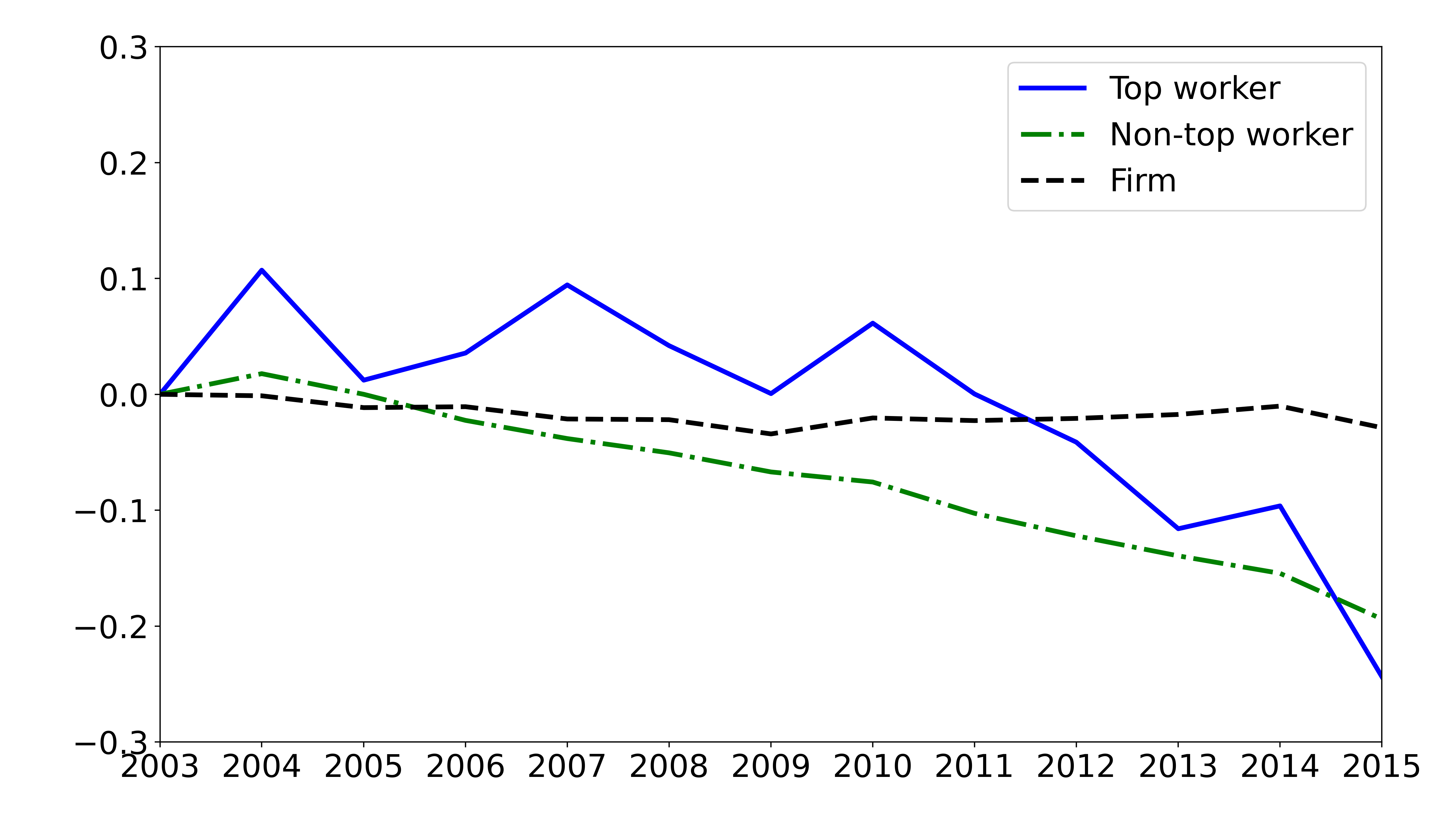}
\caption{\label{fig:orgfb2c7a7}Aggregate fixed effects}
\end{figure}

The aggregate firm fixed effect also fell slightly overall, which is due to the first sub-period 2003-2009. In the second sub-period 2009-2015, the aggregate firm fixed effect rose slightly, as can be seen in Figure \ref{fig:orgfb2c7a7}. This means that the firm fixed effect explains more of the variation in worker's labor income after 2009.

\section{Data for production function estimation}
\label{sec:orgbc6a7cd}
National Accounts Longitudinal Microdata File (NALMF) 2000-2015 is the main data source for output, capital stock, and employment. This data provides information of income statement and balance sheet of Canadian enterprises. Variables:

\begin{itemize}
\item Output: value added = Total sales of goods and services (Item 8089) - Cost of materials (Item 8320). We deflate value added by the GDP implicit price.
\item Intermediate input: Cost of materials, deflated by the price index of raw materials.
\item Capital stock: Total tangible capital assets (Item 2008), deflated by investment price index.
\item Labor input: head count of employees.
\end{itemize}

The majority of enterprises are small. Measures of capital and investment are missing for many enterprises, in particular small ones. We thus use the book value of capital stock, and we also exclude intangible assets. Whenever it is appropriate, we impute the missing values of capital stock using the law of motion of capital. We screen the data as follows:
\begin{itemize}
\item Drop a firm (for all years) if it has fewer than 2 non-top workers in more than two-thirds of sample years. The large majority of firms have only one top worker. Thus, we mostly keep firms that have at least 3 employees.
\item Drop a firm/year observation if the value of any variable (output, inputs) is extremely small (smaller than \$10 in 2002 prices).
\end{itemize}
\section{Matching under Cobb-Douglas production function}
\label{sec:org374f69b}
With the CES form of "output" function of matching, the elasticity of substitution between two worker types is not identified. If this elasticity approaches to 1, the production function takes the Cobb-Douglas form as follows:

\[f(\omega,x,y,l) = e^{\omega} (e^{\omega_x})^{\theta\alpha_x}  x^{\theta\alpha_x} y^{\theta\alpha_y} [l(x)]^{\alpha_l}.\]

We re-write it as
\[f(\omega,x,y,l) = e^{\eta} x^{\beta_x} y^{\beta_y} [l(x)]^{\alpha_l}.\]

The match efficiency cannot be identified, as both \(\omega\) and \(\omega_x\) are unobserved. Thus \(\eta\) is the Hicks-neutral technology, and it may also affect the optimal matching between \(x\) and \(y\) which was not the case with the CES form of production function.
\subsection{Matching under Cobb-Douglas production function}
\label{sec:org031d76a}
The only difference from the baseline model is that the production function is Cobb-Douglas, while other assumptions are same as before. The first-order optimal conditions for the CES case still apply to the Cobb-Douglas case. We now derive the optimal matching function. To make steps clear, we repeat all steps as before.

If matching is positively assortative, the first-order necessary conditions are given by
\[f_x - w'(x) l(x) = 0; \quad f_l - w(x) =0; \quad T'(x) = \frac{\mathcal{H}(x)}{l(x)},\]
where \(\mathcal{H}(x) = g(x) / h(T(x)) = C \cdot \frac{[T(x)]^{\lambda_y+1}}{x^{\lambda_x+1}}\) with \(C = \frac{\lambda_x \underline{x}^{\lambda_x}}{\lambda_y \underline{y}^{\lambda_y}}\).
Taking the total differentiation of \(f_l - w(x) = 0\) with respect to \(x\), it leads to
\[f_{lx} + f_{ly} \cdot T'(x) + f_{ll} l'(x) = f_x / l(x),\]
where we have applied the condition \(w'(x) = f_x / l(x)\) on the right-hand side. Given the form of production function, we notice that \(f_{lx} = \frac{\alpha_l f_x}{l(x)}\), \(f_{ly} = \frac{\alpha_l f_y}{l(x)}\), and \(f_{ll} = \frac{(\alpha_l-1) f_l}{l(x)}\). We then have
\[(\alpha_l-1) f_x + \alpha_l f_y T'(x) = (1-\alpha_l) f_l l'(x).\]

Applying to the equation above the expressions for marginal products of respectively \(x, y\) and \(l\), we obtain

\begin{equation}
\label{eq:TXCD}
(\alpha_l-1) \beta_x \frac{1}{x} + \alpha_l \beta_y \frac{1}{y} T'(x) = (1-\alpha_l) \alpha_l \frac{l'(x)}{l(x)}.
\end{equation}

We now obtain \(\frac{l'(x)}{l(x)}\). Total differentiating the optimal condition with PAM, \(T'(x) l(x) = \mathcal{H}(x)\), with respect to \(x\), we obtain \(\frac{l'(x)}{l(x)} = \frac{\mathcal{H}'(x)}{\mathcal{H}(x)} - \frac{T''(x)}{T'(x)}\). Given the Pareto distributions of worker types, \(\frac{\mathcal{H}'(x)}{\mathcal{H}(x)} = (\lambda_y+1)\frac{T'(x)}{T(x)} - (\lambda_x+1)\frac{1}{x}\). Thus,
\[\frac{l'(x)}{l(x)} = (\lambda_y+1)\frac{T'(x)}{T(x)} - (\lambda_x+1)\frac{1}{x} - \frac{T''(x)}{T'(x)}.\]

Note again, this expression for \(\frac{l'(x)}{l(x)}\) is the same as that when the optimal condition is \(T'(x) l(x) = -\mathcal{H}(x)\) (under NAM).

Thus, Equation (\ref{eq:TXCD}) becomes
\[(\alpha_l-1) \beta_x \frac{1}{x} + \alpha_l \beta_y \frac{T'(x)}{T(x)}= (1-\alpha_l) \alpha_l \left[(\lambda_y+1)\frac{T'(x)}{T(x)} - (\lambda_x+1)\frac{1}{x} - \frac{T''(x)}{T'(x)}\right].\]

Re-arranging the terms, we obtain
\[(1-\alpha_l)\left[\alpha_l(\lambda_x+1) - \beta_x\right] \frac{1}{x} + \left[\alpha_l \beta_y - (1-\alpha_l) \alpha_l (\lambda_y+1)\right] \frac{T'(x)}{T(x)} + (1-\alpha_l) \alpha_l\frac{T''(x)}{T'(x)}=0,\]
which a second-order differential equation. Guess \(y = T(x) = A x^B\) where \(A\) and \(B\) are parameters with unknown values. Using this guessed functional form, the equation above becomes

\[(1-\alpha_l)\left[\alpha_l(\lambda_x+1) - \beta_x\right] \frac{1}{x} + \left[\alpha_l \beta_y - (1-\alpha_l) \alpha_l (\lambda_y+1)\right] \frac{B}{x} + (1-\alpha_l) \alpha_l\frac{B-1}{x}=0.\]

Variable \(x\) is canceled out. Solving the equation above, we obtain the value of \(B\), as
\begin{equation}
\label{eq:B_CD}
B = \frac{(1-\alpha_l)(\alpha_l \lambda_x - \beta_x)}{\alpha_l \left[(1-\alpha_l) \lambda_y-\beta_y\right]}.
\end{equation}

Parameter \(A\) is canceled out too, we cannot find its value. Now we try another way of solving the model.

We start with the following optimal conditions:
\[f_x - w'(x) l(x) = 0; \quad f_l - w(x) =0.\]

Apply the marginal products of \(x\) and \(l\) to the first condition, we obtain \(\frac{\beta_x}{\alpha_l} \cdot \frac{f_l}{x} = w'(x)\). In this equation, substitute the second condition for \(f_l\), we get
\[w'(x) = \frac{\beta_x}{\alpha_l} \cdot \frac{w(x)}{x}.\]

This derivative suggests that \(w(x) = x^{\frac{\beta_x}{\alpha_l}}\) plus a constant, and we set the constant to zero since it is just a shifter of wages. Plug the wage equation to the second condition \(f_l - w(x) =0\) and solve for \(l(x)\), we have
\[l(x) = \left(\alpha_l e^{\eta} x^{\beta_x - \beta_x/\alpha_l} y^{\beta_y}\right)^{\frac{1}{1-\alpha_l}}.\]

Now, Substitute the expression above for \(l(x)\) in the third optimal condition \(T'(x) = \frac{\mathcal{H}(x)}{l(x)}\), it becomes
\[T'(x) = C \cdot \frac{y^{\lambda_y+1}}{x^{\lambda_x+1}} \cdot \left(\alpha_l e^{\eta} x^{\beta_x - \beta_x/\alpha_l} y^{\beta_y}\right)^{\frac{1}{\alpha_l-1}}.\]

Now guess \(y = T(x) = A x^B\). The condition above then becomes
\[ABx^{B-1} = C \cdot \frac{(Ax^B)^{\lambda_y+1}}{x^{\lambda_x+1}} \cdot \left(\alpha_l e^{\eta} x^{\beta_x - \beta_x/\alpha_l} (Ax^B)^{\beta_y}\right)^{\frac{1}{\alpha_l-1}}.\]

It can be re-written as
\[ABx^{B-1} = C \cdot A^{\lambda_y+1} \cdot \left(\alpha_l e^{\eta} A^{\beta_y}\right)^{\frac{1}{\alpha_l-1}} \cdot x^{\Delta},\]
with \(\Delta = B(\lambda_y+1) - \lambda_x-1 + \frac{\beta_x}{\alpha_l} + \frac{B\beta_y}{\alpha_l-1}\). For the equation in the above line to hold, the following identities must hold
\[B-1 = \Delta,\]

and
\[AB = C \cdot A^{\lambda_y+1} \cdot \left(\alpha_l e^{\eta} A^{\beta_y}\right)^{\frac{1}{\alpha_l-1}}.\]

From the first identity \(B-1 = \Delta\), we can solve for \(B\). It turns out that the expression for \(B\) is identical to Equation \ref{eq:B_CD}. Given the value of \(B\), we can solve the second identity for \(A\), which is
\[A = \left(B C^{-1}\right)^{\frac{1-\alpha_l}{(1-\alpha_l)\lambda_y - \beta_y}} \left(\alpha_l e^{\eta}\right)^{\frac{1}{(1-\alpha_l)\lambda_y-\beta_y}}.\]

With this, we have found the analytical form of the optimal matching function.

Positive assortative matching between top workers and non-top workers requires the numerator and denominator in the expression for \(B\) to have the same sign. Assume that the output elasticity of labor quantity is smaller than 1, then PAM requires \(\frac{\beta_x}{\lambda_x} < \alpha_l < \left(1-\frac{\beta_y}{\lambda_y}\right)\) or \(\left(1-\frac{\beta_y}{\lambda_y}\right) < \alpha_l < \frac{\beta_x}{\lambda_x}\).

Focusing on first the condition \(\frac{\beta_x}{\lambda_x} < \alpha_l < \left(1-\frac{\beta_y}{\lambda_y}\right)\), it says that the output elasticity of labor quantity is between the output elasticity of quality of two types of workers (relative to their Pareto coefficients). If this condition is satisfied, we also have \((1-\alpha_l)\lambda_y - \beta_y>0\). Not only there exists positive assortative matching, but also the higher the level of Hicks-neutral technology the higher the top-worker quality or the lower the non-top-worker quality. Thus, if the Hicks-neutral technology improves, the positivity of matching becomes weaker. The condition for positive matching implies that \(\frac{\beta_x}{\lambda_x}+ \frac{\beta_y}{\lambda_y} < 1\), which says that the sum of output elasticity for two worker types (relative to their Pareto coefficients) is (well) smaller than 1.

Instead, if condition \(\left(1-\frac{\beta_y}{\lambda_y}\right) < \alpha_l < \frac{\beta_x}{\lambda_x}\) is satisfied, the optimal matching is still positive. In this case, \(\frac{\beta_x}{\lambda_x}+ \frac{\beta_y}{\lambda_y} > 1\). Now, the higher the level of Hicks-neutral technology, the lower the quality of the top worker or the higher the quality of the non-top workers. This means that, if the Hicks-neutral technology improves, the positivity of matching becomes stronger.
\subsection{Productivity measures with Cobb-Douglas production function}
\label{sec:org31af2ed}
We first estimate the Cobb-Douglas production function below, in logarithm:
\[\ln f_{jt} = \beta_0 + \beta_x \ln x_{jt} + \beta_y \ln y_{jt} + \alpha_l \ln l_{jt} + \alpha_k \ln k_{jt} + \eta_{jt} + \varepsilon_{jt},\]
where we assume that the Hicks-neutral technology \(\eta_{jt}\) follows an AR(1) process. We implement the same procedure of estimation as in the case of CES "output" function. The only difference is that, here, instrument variables are the first lags of employment, the quality of top worker and non-top worker, intermediate inputs, and \(\widehat{\phi}\).

The measured total factor productivity for each firm is defined as
\[z_{jt} = \eta_{jt} + \widehat{\beta}_x x_{jt} + \widehat{\beta}_y \ln y_{jt},\]
which still consists of two components: the Hicks-neutral technology (total factor productivity) and the contribution of worker quality. Here, we note that the Hicks-neutral technology also measures the match efficiency, while in the CES case the match efficiency is estimated separately from Hicks-neutral technology. Thus, here we cannot examine the role of match efficiency in the productivity dynamics. Further, the definition \(z_{jt}\) for measured productivity cannot cleanly separate Hicks-neutral technology from the contribution of worker quality because the matching between \(x_{jt}\) and \(y_{jt}\) depends on Hicks-neutral technology. We could estimate the matching function and use the estimates in the productivity definition \(z_{jt}\), which is left for future study.

The measured aggregate total factor productivity is defined as \(e^{z_t} = e^{\eta_t} \cdot e^{\tilde{x}_t} \cdot e^{\tilde{y}_t}\), each component is the geometric mean of firm-level values. In logarithm, it is the weighted sum of firm-level productivity, \(z_t = \eta_t + \tilde{x}_t + \tilde{y}_t\), where \(\eta_t = \sum_{j}^{n_{t}} s_{jt}\eta_{jt}\), \(\tilde{x}_t= \sum_{j}^{n_{t}} s_{jt} \widehat{\beta}_x \ln x_{jt}\), and \(\tilde{y}_t= \sum_{j}^{n_{t}} s_{jt} \widehat{\beta}_y \ln y_{jt}\). Here, \(n_t\) is the number of firms in year \(t\), and weight \(s_{jt}\) is firm \(j\)'s share of current-price output (value added) in the current-price aggregate output. We note that \(\widehat{\beta}_x\) and \(\widehat{\beta}_y\) are sector specific.
\subsection{Productivity dynamics with Cobb-Douglas production function}
\label{sec:org060b3b1}
\begin{figure}[htbp]
\centering
\includegraphics[angle=0,scale=0.30]{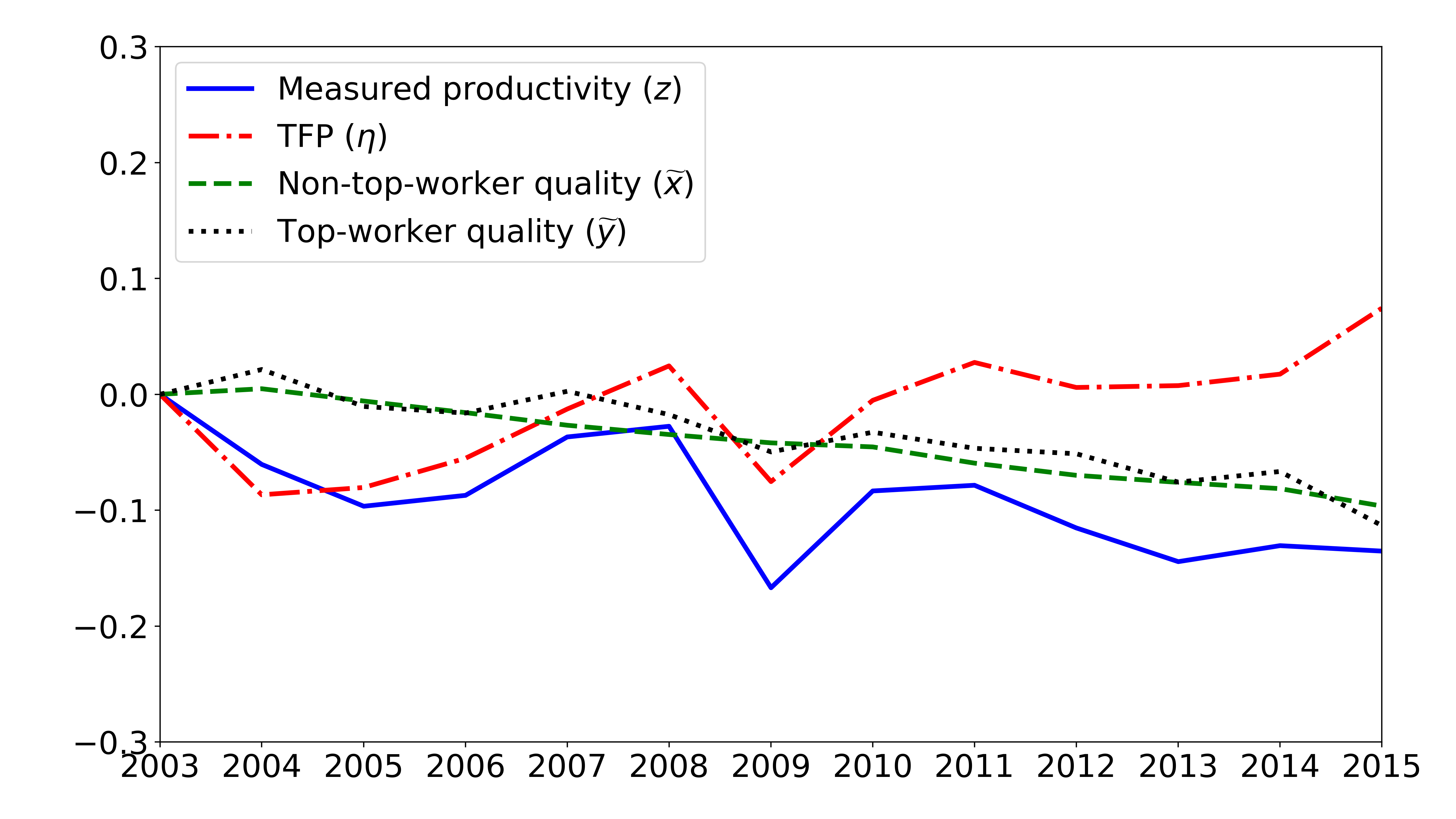}
\caption{\label{fig:orgbcb1bce}Aggregate productivity under Cobb-Douglas production function}
\end{figure}

As show in Figure \ref{fig:orgbcb1bce}, the measured aggregate total factor productivity exhibits a similar pattern of secular declines as that based on estimates in the case of CES "output" function. Over the period 2003-2015, measured aggregate total factor productivity declined on average 1.1 percent per year, while the Hicks-neutral technology improved 0.6 percent per year. The rate of Hicks-neutral technology improvement is lower than that estimated in the CES case. One source of this difference is that the Hicks-neutral technology in the Cobb-Douglas case also contains the match efficiency which declined over the sample period.

The declines of measured productivity are fully accounted for by the declines of the worker quality. On average, the component the contribution of worker quality (\(\tilde{x}+\tilde{y}\)) declined 1.7 percent per year. The contribution of top-worker quality to the measured productivity was slightly larger (more negative) than that of non-top-worker quality. Further, the declines of measured productivity are entirely due to reallocation (covariance between output shares and firm-level measured productivity), while the unweighted average of measured productivity increased.

In summary, the estimation under Cobb-Douglass "output" function shows the trends of measured productivity that display very similar patterns as those obtained in the CES case: The slowdown of measured productivity is due to the declines of worker quality; And cross-firm reallocation plays a dominant role in the declines of measured productivity.

\clearpage
\end{document}